\documentclass[amsmath,amssymb,aps,prb,12pt,superscriptaddress,longbibliography, reprint, floatfix]{revtex4-2}

\usepackage[utf8]{inputenc}
\usepackage{siunitx}
\usepackage{graphicx}
\usepackage{ragged2e}
\newcommand{\captext}[1]{\footnotesize\justifying #1}

\usepackage{amsfonts}
\usepackage{amsmath,amssymb}
\usepackage{amsthm}
\usepackage{dsfont}
\usepackage[normalem]{ulem}
\usepackage{xcolor}
\usepackage{tikz}
\usepackage{dcolumn}
\usepackage{bm}
\usepackage{changepage}      
\usepackage{etoolbox}
\usepackage[utf8]{inputenc}
\usepackage{siunitx}
\usepackage{graphicx}
\usepackage{ragged2e}

\usepackage{amsfonts}
\usepackage{amsmath,amssymb}
\usepackage{amsthm}
\usepackage{dsfont}
\usepackage[normalem]{ulem}
\usepackage{xcolor}
\usepackage{tikz}
\usepackage{dcolumn}
\usepackage{bm}
\usepackage{changepage}
\usepackage[most]{tcolorbox}
\usepackage{placeins}
\usepackage{tabularx}
\usepackage{listings}
\usepackage{soul}
\usepackage{titlesec}
\setcitestyle{super}
\titleformat{\section}
  {\bfseries\fontsize{12pt}{14pt}\selectfont}
  {}
  {0pt}
  {}
\titlespacing*{\section}{0em}{0.5em}{0.3em}

\newcommand{\I}{1\!\!1}

\newcommand{\bx}{\mathbf{x}}

\newcommand{\ba}{\mathbf{a}}
\newcommand{\bb}{\mathbf{b}}

\newcommand{\smallcirc}{\mathbin{\text{\raisebox{0.2ex}{$\scriptstyle\circ$}}}}

\definecolor{backcolor}{rgb}{0.95,0.95,0.92}
\lstdefinestyle{codingbox}{
    backgroundcolor=\color{backcolor},
    basicstyle=\ttfamily\footnotesize,
    breakatwhitespace=false,
    breaklines=true,
    captionpos=b,
    keepspaces=true,
    showspaces=false,
    showstringspaces=false,
    showtabs=false,
    tabsize=2
}
\usepackage{hyperref}
\usepackage{pdfpages}
\usepackage{pgffor}

\makeatletter
\AtBeginDocument{\let\LS@rot\@undefined}
\makeatother

\begin{document}

\title{Universality of physical neural networks
with multivariate nonlinearity}

\author{Benjamin Savinson}
 \affiliation{Optical Materials Engineering Laboratory, Department of Mechanical and Process Engineering,\\ ETH Zurich, 8092 Zurich, Switzerland}
 \affiliation{ETH AI Center, ETH Zurich, 8092 Zurich, Switzerland}
 \affiliation{Seminar for Applied Mathematics, Department of Mathematics, ETH Zurich, 8092 Zurich, Switzerland}

\author{David J. Norris}
 \email{dnorris@ethz.ch}
 \affiliation{Optical Materials Engineering Laboratory, Department of Mechanical and Process Engineering,\\ ETH Zurich, 8092 Zurich, Switzerland}
 \affiliation{ETH AI Center, ETH Zurich, 8092 Zurich, Switzerland}

 \author{Siddhartha Mishra}
 \email{siddhartha.mishra@sam.math.ethz.ch}
 \affiliation{ETH AI Center, ETH Zurich, 8092 Zurich, Switzerland}
 \affiliation{Seminar for Applied Mathematics, Department of Mathematics, ETH Zurich, 8092 Zurich, Switzerland}

 \author{Samuel Lanthaler}
 \email{samuel.lanthaler@univie.ac.at}
 \affiliation{Faculty of Mathematics, University of Vienna, 1090 Vienna, Austria}


\begin{abstract}
\noindent The enormous energy demand\cite{deVries2023} of artificial intelligence is driving the development of alternative hardware for deep learning\cite{marković2020,wetzstein2020,shastri2021}. Physical neural networks try to exploit physical systems to perform machine learning more efficiently\cite{momeni2024}. In particular, optical systems can calculate with light using negligible energy\cite{psaltis1985,lin2018,wright2022}. While their computational capabilities were long limited by the linearity of optical materials\cite{mcmahon2023}, nonlinear computations have recently been demonstrated through modified input encoding\cite{eliezer2023,wanjura2024,xia2024,yildirim2024}. Despite this breakthrough, our inability to determine if physical neural networks can learn arbitrary relationships between data---a key requirement for deep learning known as universality\cite{hornik1989}---hinders further progress\cite{mcmahon2024N&V}. Here we present a fundamental theorem that establishes a universality condition for physical neural networks. It provides a powerful mathematical criterion that imposes device constraints, detailing how inputs should be encoded in the tunable parameters of the physical system. Based on this result, we propose a scalable architecture using free-space optics that is provably universal and achieves high accuracy on image classification tasks. Further, by combining the theorem with temporal multiplexing, we present a route to potentially huge effective system sizes in highly practical but poorly scalable\cite{mcmahon2023} on-chip photonic devices. Our theorem and scaling methods apply beyond optical systems and inform the design of a wide class of universal, energy-efficient physical neural networks, justifying further efforts in their development.
\end{abstract}

\maketitle

\noindent Optical computing can utilize the low-loss and highly parallel nature of light\cite{athale2016}. By combining these properties with design and fabrication advances, photonic devices that perform differentiation\cite{silva2014,cordaro2019,zangeneh2021}, solve equations\cite{mohammadi2019,cordaro2023}, and execute matrix--vector multiplication\cite{yang2013,nikkhah2024} have been demonstrated. Because the latter operation is central to deep-learning architectures\cite{brunton2022}, increasingly sophisticated photonic neural networks have appeared\cite{shen2017,lin2018,wright2022,pai2023, hua2025, yu2025}. They belong to a broader set of physical neural networks (PNNs), which execute machine-learning tasks using the properties of physical systems\cite{marković2020,zangeneh2021}.\\
\indent To learn complex relationships between data, PNNs must include nonlinear activation functions between network layers \cite{brunton2022}. The weak nonlinearities of optical systems have therefore limited their capabilities\cite{mcmahon2023}. While nonlinear optical materials can be used\cite{wright2022}, they require high light intensities, reducing efficiency. Nonlinearities can also be included by adding electronic components to photonic integrated circuits\cite{shen2017}. This approach can be efficient\cite{bandyopadhyay2024}, but increases device complexity and lacks scalability \cite{mcmahon2023}.\\
\indent Recently, an alternative route to nonlinearity (known as structural nonlinearity), which allows both efficiency and scalability, has been proposed \cite{eliezer2023,wanjura2024,xia2024,yildirim2024}. Instead of encoding the input signal on an optical beam that enters the system, the input is encoded on tunable system parameters. A nonlinear relationship between these parameters and an output beam can be achieved when the system is optically probed. \\
\indent However, PNNs with structural nonlinearity do not map onto conventional neural networks. Thus, we do not know if such systems (called broken-isomorphism PNNs \cite{momeni2024}) can perform advanced machine-learning tasks. The deep-learning revolution is founded on a universality theorem for artificial neural networks (ANNs)\cite{hornik1989}, which proves that they have the expressive power to learn arbitrary relationships between data. Because no analogous theorem has been available to guide the development of broken-isomorphism PNNs, it is unknown which designs (if any) are universal.\\
\indent Here, we derive a universality theorem for broken-isomorphism PNNs. It provides a mathematically sharp criterion that can verify if a given system, in the limit of large size, is universal. This establishes that PNNs can in principle address even the most demanding machine-learning tasks. We then propose a scalable, free-space optical system that is provably universal, illustrating how the theorem can guide the design of useful architectures. Further, by combining insights gained from the universality criterion with temporal multiplexing, we provide a route to achieving huge effective sizes in systems that would otherwise suffer from poor scalability (for example photonic integrated circuits).

\begin{figure*}[t]
  \centering
  \includegraphics[width=14.0cm]{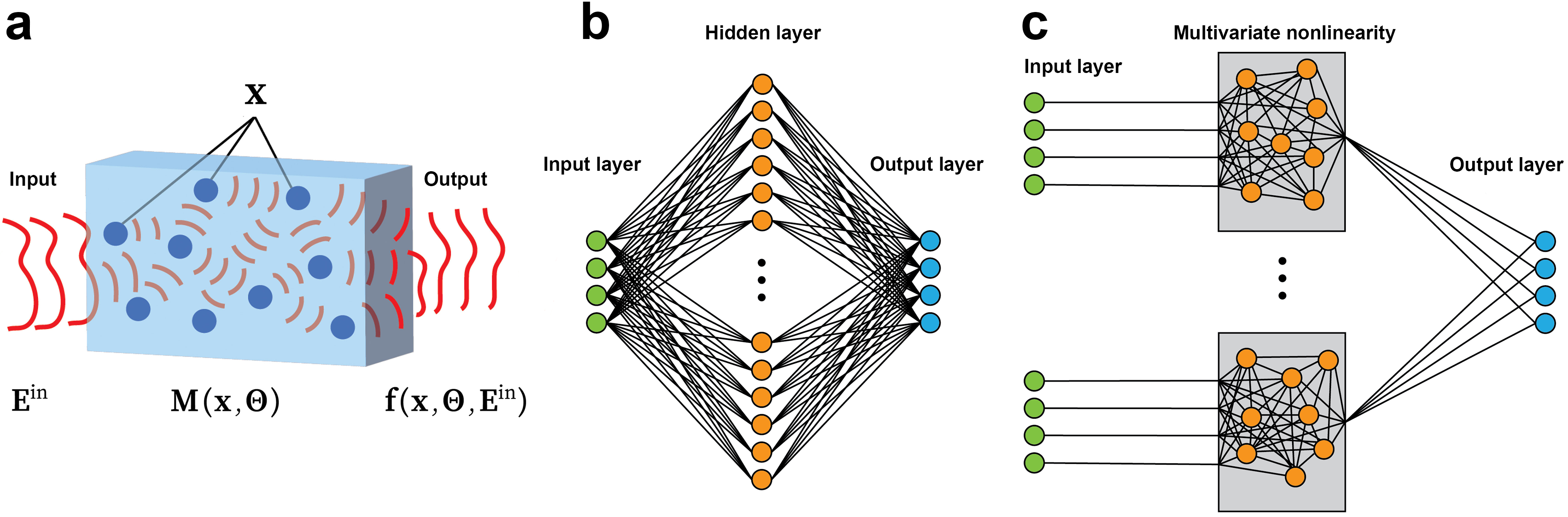}
  \caption{\captext{\textbf{Multivariate nonlinearity.} \textbf{a}, Structural nonlinearity encodes the input signal \textbf{x} on system parameters. When the system is probed by a beam $\textbf{E}^{\text{in}}$, the output \textbf{f} can be a nonlinear function of \textbf{x}. \textbf{b}, Conventional artificial neural network (ANN), with an input layer, hidden layer, and output layer. The linear matrix--vector operations mix the input components at each node, while the nonlinearity is applied element-wise between the network layers. The ANN is universal when the hidden layer is infinitely wide\cite{hornik1989}. \textbf{c}, A multivariate physical neural network (mPNN) that satisfies equation \eqref{eq:PNN_map} and criterion \eqref{eq:criterion}. Each gray box represents the multivariate, nonlinear encoding function, $\sigma_j$, which both introduces the nonlinearity and mixes the input components. The mPNN is universal when $r$, the number of copies of the input (gray boxes), goes to infinity.}}
  \label{fig:mPNNs}
\end{figure*}

\section{Universality theorem}
\label{sec: I}
\noindent The interaction of an optical wave with a linear physical system can be described by a transfer matrix \textbf{M}, which depends on the physical parameters of the system. By representing the incoming wave as a vector $\textbf{x}$ of dimension $d$, the output is given by the linear relationship $\textbf{y} = \textbf{M}\textbf{x}$. A PNN based on such a system can perform linear classification tasks on input $\textbf{x}$ if \textbf{M} depends on a set of trainable parameters $\boldsymbol{\Theta}$, giving output 
\begin{equation}
    \textbf{f}(\textbf{x},\boldsymbol{\Theta}) :=  \textbf{M}(\boldsymbol{\Theta})\; \textbf{x}~.
\end{equation}
\indent To include structural nonlinearity, the input $\textbf{x}$ is encoded on system parameters rather than the incoming wave\cite{eliezer2023,wanjura2024,xia2024,yildirim2024} (Fig. \ref{fig:mPNNs}a). The system is probed with beam $\textbf{E}^{\text{in}}$, yielding
\begin{equation}
    \textbf{f}(\textbf{x},\boldsymbol{\Theta}, \textbf{E}^{\text{in}}) := \textbf{M}(\textbf{x},\boldsymbol{\Theta}) \;\textbf{E}^{\text{in}}~.
\label{eq: lin_system}
\end{equation}
\noindent This output can depend nonlinearly on \textbf{x}, for example, if the probe beam interacts multiple times with the system (known as input replication)\cite{eliezer2023,wanjura2024,xia2024,yildirim2024}.\\
\indent Inspired by the above system, we introduce a general mathematical framework for such PNNs. Specifically, we consider systems that can be mapped onto the form
\begin{align}
\label{eq:PNN_map}
\textbf{f}(\textbf{x},\textbf{a}_j,\textbf{b}_j, \textbf{c}_j) = \sum_{j=1}^r \textbf{c}_j 
\sigma(\textbf{a}_j \! \smallcirc \! \textbf{x} + \textbf{b}_j)~.
\end{align}
\noindent Here, $\textbf{f}$ is the output vector, $\textbf{c}_j$ is a vector of trainable weighting factors, and $\sigma_j$ is a multivariate, nonlinear encoding function that depends on input $\textbf{x}$. Vectors 
$\textbf{a}_j$ and $\textbf{b}_j$ are trainable parameters that permit efficient rescaling of $\textbf{x}$ during pre-processing ($\textbf{a}_j \! \smallcirc\! \textbf{x}$ is an element-wise vector product). Physically, equation \eqref{eq:PNN_map} describes a PNN in which $r$ copies of the input are nonlinearly encoded on different (for example, spatially divided) sections of the system, represented by $\sigma$. The responses from the individual sections are then linearly combined with weights $\textbf{c}_j$ to yield the system output.\\
\indent In Supplementary Information Section S1, we derive a universality theorem for systems of the form \eqref{eq:PNN_map}. It requires that the encoding functions $\sigma$ satisfy a strict criterion. Namely, for $r \to \infty$, the system is universal if and only if no set of integers $n_i$ exists for which
\begin{align}
\label{eq:criterion}
\frac{\partial^{n} \sigma(\bx)}{\partial x_1^{n_1} \dots \partial x_d^{n_d}} \equiv 0~,
\end{align}
where $n=n_1 + \dots + n_d$. This means that the functions $\sigma$ must contain arbitrary coupling orders between all input components $x_i$. This constraint is stronger than simply requiring that the encoding function $\sigma$ is non-polynomial. For example, $\sigma(\textbf{x}) = \sum_{k=1}^d  \exp{ix_k}$ does not mix different input components, and equation \eqref{eq:criterion} is not satisfied [for $l\neq m$, $\partial^2\sigma(\bx) / \partial x_l \partial x_m = 0$]. In contrast, $\sigma(\bx) = \exp\bigl( \sum_{k=1}^d ix_k \bigr)$  provides coupling to arbitrary order and fulfills the universality criterion. We note that universality holds similarly for pairwise different encoding functions $\sigma_j$ that fulfill equation \eqref{eq:criterion}.\\
\indent At first glance, equation \eqref{eq:PNN_map} may appear to describe a conventional ANN with one hidden layer (Fig. \ref{fig:mPNNs}b). However, in ANNs the linear matrix--vector operations mix the input components, while the nonlinearity is applied element-wise between the network layers. This is fundamentally different to our PNNs (Fig. \ref{fig:mPNNs}c), where the linear operations $\textbf{a}_j\! \smallcirc\!  \textbf{x} + \textbf{b}_j$ are performed element-wise, while the multivariate function $\sigma_j$ both introduces the nonlinearity and mixes the input components. This `multivariate nonlinearity' represents a distinct (and unexplored) deep-learning paradigm. It occurs naturally in physical systems that map onto equation \eqref{eq:PNN_map}, which we call multivariate PNNs (or mPNNs).

\begin{figure*}[t]             
  \centering
  \includegraphics[width=15cm]{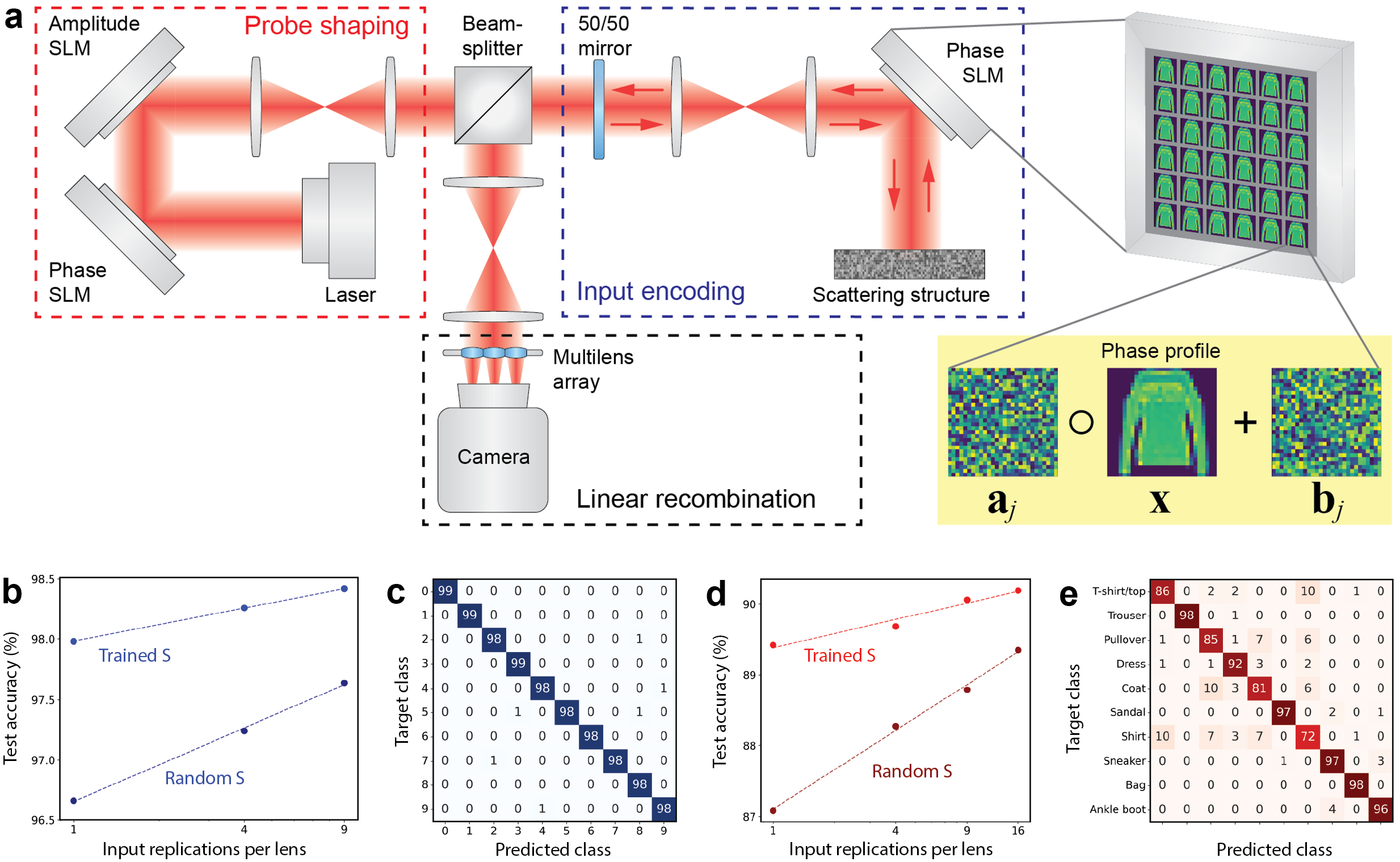}
  \caption{\captext{\textbf{Proposed physical neural network with multivariate nonlinearity.} \textbf{a}, A free-space optical system consisting of three blocks. The beam-shaping block spatially prepares the phase and amplitude of a laser beam with two spatial light modulators (SLMs). The beam then enters the input-encoding block, which includes a mirror with 50\% reflectivity, a phase SLM that contains multiple copies of the input, and a scattering structure. Due to the mirror, the beam interacts multiple times with the SLM and the scattering structure, leading to multivariate nonlinear encoding. The linear-recombination block contains a multilens array in front of an imaging camera, which recombines the different spatial components of the beam to yield the output. Lenses are also placed in the system to route and collimate the beam. \textbf{b}, \textbf{d}, Best testing accuracy achieved for the MNIST and Fashion-MNIST datasets, respectively, for varying numbers of input copies per lens. We observe clear scaling in the number of input copies for both trainable and random $\textbf{S}$. \textbf{c}, \textbf{e} The confusion matrix of the best-performing model with trainable $\textbf{S}$ on the MNIST and Fashion-MNIST testing sets, respectively.}}
  \label{fig:free_space}
\end{figure*}
\section{Free-space implementation}
\label{sec:free-space}
\noindent To show the utility of our theorem, we exploit it to design a universal mPNN. Although many physical systems could be considered, we focus on free-space optics due to its advantageous scaling\cite{mcmahon2023}. Our proposed setup (Fig. \ref{fig:free_space}a) consists of a light source (laser), a sequence of optical elements, and a detector (camera). It is organized into three blocks, as prescribed by equation \eqref{eq:PNN_map}. The first prepares the probe beam by locally tailoring its phase and amplitude with two spatial light modulators (SLMs), according to the weighting factors $\textbf{c}_j$. The probe then interacts with the $r$ input copies in the second block, according to a multivariate function $\sigma$. The third block recombines the beam components, according to the summation.\\
\indent The input encoding is performed by a 50\% reflective mirror, an SLM, and a scattering structure. After the beam passes the mirror, its phase is modified by the SLM, which contains $r$ copies of the input $\tilde{\textbf{x}}= \textbf{a}_j \!\smallcirc \textbf{x} + \textbf{b}_j$ as a phase profile in its pixels (Fig. \ref{fig:free_space}a). The impact of the SLM can be represented by a diagonal matrix $\textbf{T}(\tilde{\textbf{x}})$. The beam then propagates to the scattering structure and returns to the SLM. For simplicity, we describe this entire step with a single matrix $\textbf{S}$, assumed to be symmetric and unitary (as in an ideal optical system). If desired, diffraction and losses could be included explicitly with Toeplitz matrices and attenuation factors (see Supplementary Information Section S2). \\
\indent Because of the partially reflective mirror, the beam interacts multiple times with the SLM and the scattering structure. The full encoding block can be written as an input-dependent matrix $\textbf{M}(\tilde{\textbf{x}})$ given by (Supplementary Information Section S2)
\begin{align}
\label{eq:Neumann}
    \textbf{M}(\tilde{\textbf{x}})=\hspace{-0.5em}
    \underbrace{\text{r}_\text{m} \, \I}_{\text{\scriptsize\shortstack{direct\\reflection}}}
    \hspace{-0.5em}+\hspace{0.2em} 
    \underbrace{\text{t}_\text{m} \, \I \left[ \textbf{T}(\tilde{\textbf{x}}) \, \textbf{S} \, \textbf{T}(\tilde{\textbf{x}}) \, \text{t}_\text{m} \, \I \right]}_{\text{\scriptsize\shortstack{one pass through\\the encoding block}}} 
    +\ 
    \hspace{-0.8em}\underbrace{\cdots}_{\text{\scriptsize\shortstack{multiple\\passes}}}~,
\end{align}
where $\text{r}_\text{m}$ and $\text{t}_\text{m}$ are the reflection and transmission coefficients of the mirror, $\I$ is the identity matrix and the ellipses account for multiple passes through the encoding block. If $\text{r}_\text{m}$ and $\text{t}_\text{m}$ are equal, equation \eqref{eq:Neumann} reduces to a Neumann series \cite{cordaro2023}, equivalently written as
\begin{equation}
\label{eq:Neumann_inv}
\textbf{M}(\tilde{\textbf{x}}) = \text{r}_\text{m} \left[ \I - \text{r}_\text{m} \, \textbf{T}(\tilde{\textbf{x}}) \, \textbf{S} \, \textbf{T}(\tilde{\textbf{x}}) \right]^{-1}~.
\end{equation}
By tailoring the amplitude and phase of the probe beam $\textbf{E}^{\text{in}}$ in the first block, the different matrix elements of $\textbf{M}(\tilde{\textbf{x}})$ can be weighted.  Although expression \eqref{eq:Neumann_inv} is obviously nonlinear in the matrix elements, it is not clear if it provides the type of multivariate nonlinearity required to fulfill the universality criterion in equation \eqref{eq:criterion}. \\
\indent Indeed, matrices $\textbf{S}$ exist for which the criterion is not fulfilled. For example, if $\textbf{S}=\I$, we obtain a diagonal matrix \textbf{M} in which each element depends on only one input component. This yields element-wise nonlinearity, which is insufficient. Rather, $\textbf{S}$ should introduce mixing between the different beam components. Intuitively, we might expect such mixing if $\textbf{S}$ is not diagonal. In Supplementary Information Section S3, we formalize this intuition. We define an encoding function based on the matrix elements and prove that the set of symmetric, unitary matrices $\textbf{S}$ for which the universality criterion is fulfilled is dense. That is, the criterion is fulfilled for almost all such matrices.\\
\begin{figure}[t]             
  \centering
  \includegraphics[width=\columnwidth]{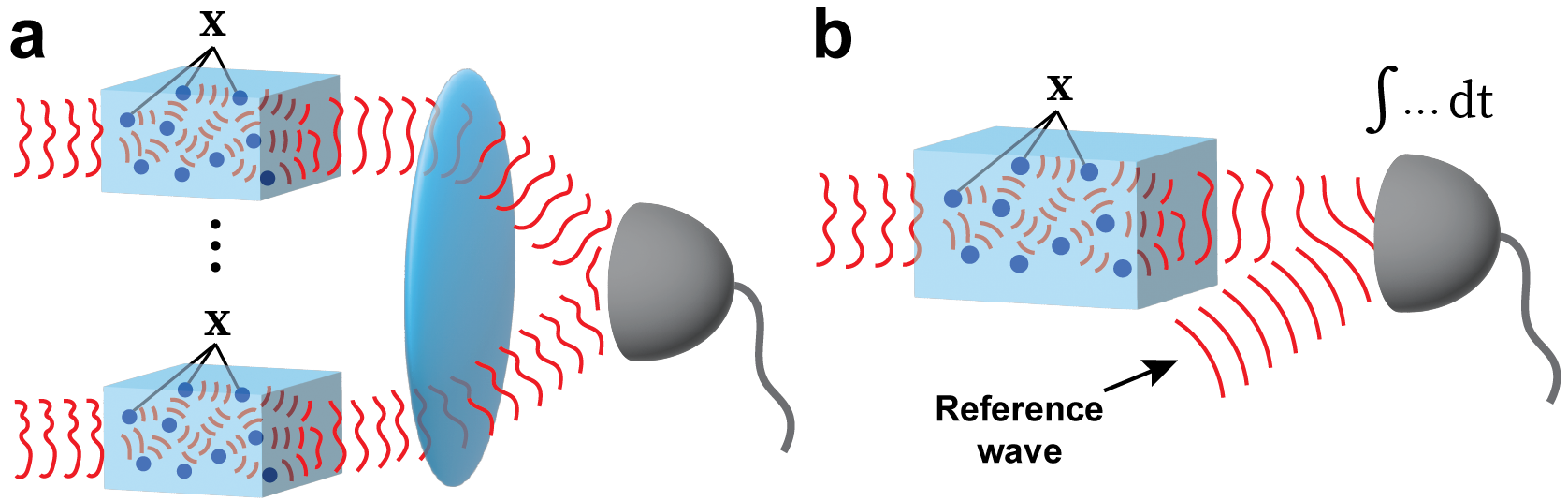}
  \caption{\captext{\textbf{Spatial and temporal scaling strategies.} \textbf{a}, System that is scaled spatially, similar to that proposed in Fig. \ref{fig:free_space}a. Multiple copies of the input \textbf{x} are projected in different spatial subsections and recombined with a lens. \textbf{b}, System that is temporally scaled. The input is projected only once in the system, but the trainable parameters are varied in time. The output is the time-integrated detector signal. A reference wave is used to retain phase information.
  }}
\label{fig:schematics}
\end{figure}
\indent In a real device, more control over $\textbf{S}$ is advantageous. This provides additional trainable parameters. The mirror reflectivity is also tunable, as the criterion is equally satisfied for the general case $\text{r}_\text{m} \neq \text{t}_\text{m}$.\\
\indent Finally, to recombine the different spatial components of the beam, the third block of our PNN passes the beam through a multilens array placed in front of an imaging detector. The lenses perform Fourier transforms, linearly combining the spatial frequencies in their specific section of the beam. In the detector plane behind each lens (Fig. \ref{fig:schematics}a), the beam profile then takes the form 
\begin{align}
    \label{eq:inout}
    \textbf{E}_k^\text{out}(\tilde{\bx}) = \mathcal{F}\left\{\text{r}_\text{m} \left( \I - \text{r}_\text{m}\textbf{T} (\tilde{\bx})\textbf{S} \textbf{T}(\tilde{\bx}) \right)^{-1} \textbf{E}^\text{in}_k\right\}~,
\end{align}
where $\mathcal{F}{\{\cdot\}}$ is the two-dimensional (2D) Fourier transform and $\textbf{E}^\text{in}_k$ denotes the beam section corresponding to the $k^{\textrm{th}}$ lens. In Supplementary Information Section S3, we prove that, for arbitrary but fixed components of $\textbf{E}_k^{\text{out}}(\tilde{\bx})$, this transformation maps the system to the form in equation \eqref{eq:PNN_map} with the output dimension equal to the number of lenses. Because criterion \eqref{eq:criterion} for the encoding function $\sigma_j$ is satisfied, the system in Fig. {\ref{fig:free_space}a is universal. The trainable parameters of the PNN are the input-rescaling parameters $\ba_j$ and $\bb_j$, the probe beam $\textbf{E}^\text{in}$, and (optionally) the scattering matrix $\textbf{S}$.
\hspace{-1.0em}
\section{Numerical experiments}
We assessed the abilities of the proposed architecture on two classification tasks: the MNIST\cite{lecun1998} and Fashion-MNIST\cite{xiao2017} datasets. Both contain 50,000 training and 10,000 grayscale testing images that we downsample to $14\times14$ pixels. In our simulations, which follow equation \eqref{eq:inout}, we assume that $\textbf{S}$ is a block-matrix, corresponding physically to input copies having some spatial separation on the SLM (see Supplementary Information Sections S3.4). This assumption is reasonable for free-space implementations, while making the numerics tractable by reducing the matrix inversion to smaller sub-blocks. The system output is an intensity measurement at 10 separate locations on the detector, each assumed to lie behind a lens performing a Fourier transform. For details on the numerical implementation and training parameters see Supplementary Information Section S4. \\
\indent We considered two possibilities for the scattering matrix $\textbf{S}$: random and fully controllable (trainable). We then trained the respective systems for varying numbers of input copies on the SLM. The results of our numerical experiments (Fig. \ref{fig:free_space}b--e) reveal that, for a trained $\textbf{S}$, the architecture achieves a maximum accuracy of $98.42\%$ on MNIST and $90.19\%$ on Fashion-MNIST. For a random $\textbf{S}$, we obtained $97.64\%$ and $89.35\%$, respectively. These results are comparable to small ANNs.\cite{wanjura2024, mcmahon2024N&V} \\
\indent Furthermore, for both MNIST and Fashion-MNIST, we observe scaling laws for the test accuracy in the number of input copies $r$ for trained and random $\textbf{S}$ (Fig. \ref{fig:free_space}b,d). This empirically supports that mPNNs scale with $r$, as predicted by the universality theorem. Further performance increases were limited by overfitting, with the training accuracy eventually reaching $100\%$ on both tasks without data augmentation. While overfitting is generally undesirable, it is consistent with a system that is highly expressive.
\section{Scaling}
\indent Our universality theorem applies to systems in which the number of copies $r \to \infty$. In experiments, two approaches to scale mPNNs can be considered. The most obvious is to increase the system size spatially. For free-space optical systems, SLMs can modulate over $10^7$ spatial pixels in parallel. Thus, for small-to-medium dimensional inputs, many copies can be projected in parallel to achieve large effective values for $r$. \\
\indent Spatial scaling is less favorable for photonic integrated circuits, where the effective system size is constrained by their planar architecture. However, this limitation can be avoided by scaling the system temporally (Fig. \ref{fig:schematics}b). Equation \eqref{eq:PNN_map} can be rewritten as a time integral
\begin{align}
    \sum_{j=1}^r \textbf{c}_j 
\sigma(\textbf{a}_j \!\smallcirc\! \textbf{x} + \textbf{b}_j) \to \int \textbf{c}_t\sigma (\textbf{a}_t\!\smallcirc\! \bx + \bb_t) dt~,
\label{eq:temporal_scaling}
\end{align}
\noindent with time-varying trainable parameters $\textbf{a}_t$, $\textbf{b}_t$, and $\textbf{c}_t$. In this continuous-time formulation, the input appears only once in the system, and the value of $r$ becomes bounded by the integration timescale and the switching speed of the parameter values. \\
\indent A complication is that this approach requires phase-sensitive detection. This can be addressed by interfering a reference wave $\textbf{E}_0$ with the output beam on the detector (Fig. \ref{fig:schematics}b). This yields an output
\begin{align}
    \textbf{f}(\bx, \ba_t, \bb_t, \mathbf{c}_t) = \int \big|\textbf{E}_0 + \mathbf{c}_t \sigma(\mathbf{a}_t \!\smallcirc\! \mathbf{x} + \mathbf{b}_t)\big|^{\smallcirc 2} \, dt,
\label{eq:temporal_detection}
\end{align}
\noindent where $|\mathbf{v}|^{\smallcirc 2} := \mathbf{v} \smallcirc \mathbf{v}^\ast$ denotes the element-wise absolute-value-squared operation applied with intensity detection. We formally prove that universality is preserved for $\textbf{E}_0 \neq 0$ in Supplementary Information Section S5. This implies that time integration of the detector while varying the system parameters can lead to universality. \\
\indent Illustratively, if switching speeds of up to 100 GHz can be achieved in integrated photonics, effective system sizes of up to $r\sim 10^7$ could result, assuming each measurement is integrated over 100~ms. If temporal and spatial scaling are combined even larger values for $r$ are possible.

\section{Conclusion}
We have derived a general theorem for multivariate physical neural networks that specifies how inputs should be encoded in physical systems to obtain universality. Guided by this result, we have designed a free-space optical system that satisfies the universality criterion. Numerical simulations show it is highly expressive, with test performance scaling with the system size. The theorem also suggests how temporal multiplexing can resolve scalability issues in spatially constrained systems, such as photonic integrated circuits.  It remains open which design for universal physical neural networks will generalize most favorably to advanced machine learning tasks. Additionally, hardware realizations of the best-performing systems must be realized. Finally, efficient training algorithms are needed\cite{momeni2024} to enable deployment on large-scale problems. While each of these steps requires significant further innovation, the universality theorem provides a rigorous theoretical foundation to justify these efforts.

\clearpage

\pagestyle{empty}
\includepdfset{fitpaper=true, pagecommand={}}


\end{document}


\title{Supplementary information for\\ ``Universality of physical neural networks with\\
multivariate nonlinearity''}
\author{\vspace{0.7em}Benjamin Savinson}
 \affiliation{Optical Materials Engineering Laboratory, \vspace{-0.7em} \\ 
 Department of Mechanical and Process Engineering, ETH Zurich, 8092 Zurich, Switzerland}
 \affiliation{ETH AI Center, ETH Zurich, 8092 Zurich, Switzerland}
 \affiliation{Seminar for Applied Mathematics, Department of Mathematics, ETH Zurich, 8092 Zurich, Switzerland}
\author{David J. Norris}
 \affiliation{Optical Materials Engineering Laboratory, \vspace{-0.7em} \\ 
 Department of Mechanical and Process Engineering, ETH Zurich, 8092 Zurich, Switzerland}
 \affiliation{ETH AI Center, ETH Zurich, 8092 Zurich, Switzerland}
 \author{Siddhartha Mishra}
 \affiliation{ETH AI Center, ETH Zurich, 8092 Zurich, Switzerland}
 \affiliation{Seminar for Applied Mathematics, Department of Mathematics, ETH Zurich, 8092 Zurich, Switzerland}
 \author{Samuel Lanthaler\vspace{0.7em}}
 \affiliation{Faculty of Mathematics, University of Vienna, 1090 Vienna, Austria}
%
%

\maketitle

\vspace{-2.8em}
\tableofcontents
\thispagestyle{empty}  
\newpage
\clearpage
\setcounter{page}{1}  

%
%
%
%
\section{Universality of physical neural networks}
\label{SIsec:Universality}
\noindent Here, we rigorously analyze the universality of physical neural networks (PNNs), specifically those that can be mapped onto an input--output relationship $\textbf{f}: \R^d \to \R^m$ of the form,
\begin{align}
\label{eq:mPNN}
\textbf{f}(\bx) = \sum_{j=1}^r \textbf{c}_j \sigma(\ba_j \smallcirc \bx + \bb_j),
\end{align}
for inputs $\bx \in \R^d$ and tunable coefficients $\textbf{c}_j\in \R^m$, $\ba_j, \bb_j \in \R^d$. $\textbf{a}_j$ and $\textbf{b}_j$ allow $\textbf{x}$ to be rescaled and biased, respectively. The summation runs over $r$ copies of the input, as discussed in the main text. The system--response function $\sigma(\bx, \ba_j, \bb_j): \R^d \to \R$, which we refer to as the \emph{multivariate nonlinear encoding function}, characterizes the input encoding in the system. 
%
%
\subsection{Distinction between ANNs and multivariate PNNs}
\noindent As mentioned in the main text, equation \eqref{eq:mPNN} may appear similar to the expression that describes a three-layer artificial neural network (ANN). However, two key differences should be noted (Table ~\ref{SItable:ANN_vs_mPNN}). First, the nonlinear encoding function ($\sigma: \R^d \to \R$) is a genuine multivariate function, which does not just act component-wise, but instead, mixes (or couples) the various input components. Second, the affine mapping $\textbf{x} \mapsto \ba_j \smallcirc \bx + \bb_j$ in our PNN operates component-wise and does not mix components ($\textbf{a}_j$ is a vector, not a full matrix).
\begin{table}[b]
\caption{\textbf{Artificial neural networks versus PNNs that map onto equation \eqref{eq:mPNN}}}
\label{SItable:ANN_vs_mPNN}
\begin{ruledtabular}
\begin{tabular}{@{}lcc@{}}
\textrm{} & \textrm{Artificial neural networks} & \textrm{Physical neural networks} \\
\colrule
\textbf{Layer}
& $\mathbf{x}\mapsto \rho(\mathbf{A}\mathbf{x}+\mathbf{b})$
& $\mathbf{x}\mapsto \sigma(\mathbf{a}_j \smallcirc \mathbf{x}+\mathbf{b}_j)$ \\
\textbf{Affine part} & Mixes components & Operates component-wise \\
\textbf{Nonlinear part} & Operates component-wise & Mixes components \\
\end{tabular}
\end{ruledtabular}
\end{table}

%
%
\subsection{The universality theorem for multivariate PNNs}

\noindent Our aim is to determine the form that $\sigma$ must have such that the PNNs described by equation \eqref{eq:mPNN} are universal. We now introduce the following notation for the partial derivatives of a mapping $\sigma: \R^d \to \R$, with multi-index $\balpha = (\alpha_1,\dots, \alpha_d) \in \N_0^d$ and $\bx \in \R^d$:
\begin{align}
\label{eq:Universality_constraint}   
\frac{\partial^{\balpha} \sigma(\bx)}{\partial \bx^{\balpha}}
=
\frac{\partial^{|\alpha|} \sigma(\bx)}{\partial x_1^{\alpha_1} \dots \partial x_d^{\alpha_d}},
\quad
|\alpha| = \alpha_1 + \dots + \alpha_d~.
\end{align}
\noindent We then obtain the following fundamental result:
\vspace{-1em} 
\begin{indentedtheorem}
\label{thm:universality}
Assume that $\sigma \in C^\infty(\R^d)$. Let $\Omega \subset \R^d$ be a compact domain. PNNs of the form \eqref{eq:mPNN} are universal in $C(\Omega; \R^m)$, if and only if $\sigma$ is non-degenerate in the following sense: no multi-index 
$\balpha = (\alpha_1,\dots, \alpha_d) \in \N_0^d$ exists such that $\partial^{\balpha} \sigma / \partial \bx^{\balpha} \equiv 0$.
\end{indentedtheorem}
\vspace{+1em} 
\noindent Before coming to the proof, we make two remarks:
\vspace{-1em} 
\begin{indentedremark}
\label{rem:universality}
More generally, it is possible that our system has a nonlinear encoding function that depends on additional controllable physical parameters, $\bm{\Theta} \in \R^m$, that is, we have a function $\sigma: \Omega \times \R^m \to \R$. In this case, the non-degeneracy condition generalizes to the requirement that, for all $\bm{\Theta}$,  
\begin{align}
\label{eq:RemarkS1.1} 
\frac{\partial^{\balpha} \sigma(\bx,\bm{\Theta})}{\partial \bx^{\balpha}} \not \equiv 0~.
\end{align}
\end{indentedremark}
\vspace{+0em} 
\vspace{-2em}
\begin{indentedremark}
Theorem \ref{thm:universality} assumes that $\sigma$ is real-valued. The proof of universality immediately extends to the case where $\sigma: \R^d \to \mathbb{C}$ is complex-valued. PNNs of the form \eqref{eq:mPNN} are still universal, provided that $\sigma$ satisfies the same non-degeneracy condition and provided that we allow the coefficients $\textbf{c}_j$ to be complex-valued. The parameters $\ba_j$ and $\bb_j$ are still assumed to be real-valued.
\end{indentedremark}
%
%
\subsection{Proof of the universality theorem for multivariate PNNs}
\noindent Our proof of the universality theorem easily follows from a reduction to the case of scalar-valued outputs, that is, reduction to the case $m=1$. To prove Theorem \ref{thm:universality}, we first recall two useful results in the scalar-valued case. The first is the following version of the Riesz Representation Theorem:
\vspace{-1em}
\begin{indentedproposition}
\label{prop:riesz}
Let $\Omega \subset \R^d$ be a compact domain.
The dual of $C(\Omega)$ is the space of finite measures on $\Omega$. More precisely, for any continuous linear functional $\ell: C(\Omega) \to \R$, there exists a finite measure $\mu$ on $\Omega$, such that
\begin{align}
\label{eq:PropositionS1.1} 
\ell(f) = \int_{\Omega} f(\bx) \, d\mu(\bx)~.
\end{align} 
\end{indentedproposition}
\vspace{+0em} 
\begin{adjustwidth}{3em}{0pt} 
\begin{proof}
This is a special case of the general version of the Riesz Representation Theorem (see Theorem 2.14 in ref. \citenum{Rudin1987}).
\end{proof}
\end{adjustwidth}
\vspace{+0.5em}
\noindent Second, we recall the following consequence of the Hahn-Banach Extension Theorem:
\vspace{-1em}
\begin{indentedproposition}
\label{prop:dense}
Let $V \subset C(\Omega)$ be a linear subspace. Then $V \subset C(\Omega)$ is not dense, if and only if there exists a finite measure $\mu$, such that $\mu \ne 0$ and 
\begin{align}
\label{eq:PropositionS1.2} 
\int_{\Omega} f(\bx) \, d\mu(\bx) = 0, \quad \forall \, f \in V~.
\end{align}
\end{indentedproposition}
\vspace{+0.5em}
\begin{adjustwidth}{3em}{0pt} 
\begin{proof}
If $V$ is not dense in $C(\Omega)$, then there exists a function $f_0\in C(\Omega) \setminus \bar{V}$. Any element $f$ in the span of $V \cup \{f_0\}$ can be uniquely written in the form $f = v + \lambda f_0$, where $v \in V$ and $\lambda \in \R$. We can define a bounded linear functional $\tilde{\ell}: C(\Omega) \to \R$ by $\tilde{\ell}(f) := \lambda$. Linearity is immediate. Boundedness follows from the observation that 
\begin{align}
\label{eq:Proof_propS1.2A} 
\Vert \tilde{\ell} \Vert 
= 
\sup_{\lambda \ne 0, v\in V} \frac{|\tilde{\ell}(v+\lambda f_0)|}{\Vert v+\lambda f_0\Vert}
= 
\sup_{v\in V} \frac{1}{\Vert \lambda^{-1} v + f_0 \Vert}~.
\end{align}
\noindent Since $V$ is a vector space, we can replace the supremum over $v \in V$ in the last expression by the supremum over $w = -\lambda^{-1}v \in V$, to obtain
\begin{align}
\label{eq:Proof_propS1.2B} 
\Vert \tilde{\ell} \Vert 
= \sup_{w\in V} \frac{1}{\Vert f_0 - w\Vert}
= \frac{1}{\inf_{w\in V} \Vert f_0 - w\Vert}~.
\end{align}
The last denominator is strictly positive, since $f_0 \not \in \bar{V}$, by assumption. By the Hahn-Banach Theorem, there exists an extension $\ell: C(\Omega) \to \R$ of $\tilde{\ell}$, with $\Vert \ell \Vert = \Vert \tilde{\ell} \Vert$. By Proposition \ref{prop:riesz}, the functional $\ell$ is represented by a finite measure $\mu$. Thus, we have shown that if $V$ is not dense in $C(\Omega)$, then there exists a non-trivial $\mu$, such that 
\begin{align}
\label{eq:Proof_propS1.2C} 
\ell(f) = \int_{\Omega} f(\bx) \, d\mu(\bx) = 0, \quad \forall \, f \in V~.
\end{align}
This confirms that if $V$ is not dense, then there exists a non-trivial functional vanishing identically on $V$, as claimed.

\indent The other direction is easy: If $\ell$ is a continuous functional on $C(\Omega)$, such that $\ell(v) = 0$ for all $v\in V$, and if $V$ is dense, then it follows by continuity that $\ell(f) = 0$ for all $f\in C(\Omega)$. Thus, $\ell = 0$. Hence, there cannot be a non-trivial functional vanishing on $V$ in this case.
\end{proof}
\end{adjustwidth}
\vspace{+0.5em}
We are now ready to proceed with the proof of Theorem \ref{thm:universality}, that is, the universality theorem for mulitvariate PNNs.
\vspace{+0.5em}
\begin{adjustwidth}{3em}{0pt} 
\begin{proof}[Proof of Theorem \ref{thm:universality}]
We start our proof by observing that it will be sufficient to prove universality for the scalar-valued case (output dimension $m=1$), since the vector-valued case is then immediate: Indeed, if $\mathbf{f}: \Omega \to \R^m$ is a vector-valued continuous function for $m>1$, then universality for the scalar-valued case implies that we can approximate each individual component $\textbf{f}_k(x)$ to any desired accuracy by a PNN with scalar coefficients $c_{j}^{(k)}$ and scaling/bias coefficients $\ba_j^{(k)}$, $\bb_j^{(k)}$, for all $k =1,\dots, m$. Choosing vector coefficients of the form $\textbf{c}_j^{(k)} = c_j^{(k)} \textbf{e}_k$, with $\textbf{e}_k$ the $k^\textrm{th}$ unit vector and summing over all $k$ and $j$ indices, that is, considering the resulting PNN of the form,
\[
\sum_{j,k} \textbf{c}_j^{(k)} \sigma\left( \ba_j^{(k)} \smallcirc \bx + \bb_j^{(k)} \right),
\]
it then follows that all components of $\textbf{f}$, and hence the vector-valued function $\textbf{f}$ itself, are approximated to the desired accuracy. Thus, if universality is established in the scalar-valued case, then universality for the vector-valued case is immediate.

Given the observation above, we focus on the scalar-valued case ($m=1$) for the remainder of this proof. The dual of the space of continuous functions on a compact domain is the space of finite measures. If the subspace of all functions of the form \eqref{eq:mPNN} were not dense in $C(\Omega)$, a non-zero measure $\mu\not \equiv 0$ on $\Omega$ would exist such that 
\begin{align}
\label{eq:ProofS1.1_eqA}  
\int_\Omega \sum_{j=1}^r c_j \sigma(\ba_j \smallcirc \bx + \bb_j) \, d\mu(\bx) = 0~,
\end{align} 
\noindent for all possible choices of $r \in \N$ and $\ba_j, \bb_j \in \R^d$, $c_j \in \R$. Our goal is to show that no such $\mu$ can exist, from which it then follows that the subspace of all functions of the form \eqref{eq:mPNN} is dense in $C(\Omega)$.
To this end, it suffices to show that the equality
\begin{align}
\label{eq:ProofS1.1_eqB} 
\int_\Omega \sigma(\ba\smallcirc\bx + \bb) \, d\mu(\bx) = 0, \quad \forall \, \ba,\bb\in \R^d~,
\end{align}
already implies that $\mu \equiv 0$~. \\
\indent For the moment, we assume a fixed multi-index $\balpha = (\alpha_1,\dots, \alpha_d)$. If $\mu$ satisfies condition \eqref{eq:ProofS1.1_eqB} for all $\ba$ and $\bb$, then given any $\bx_0\in \Omega$, we can choose $\bb = \bx_0$ and differentiate in $\ba$ to obtain (chain rule),
\begin{align}
\label{eq:ProofS1.1_eqC} 
0 
&= 
\frac{\partial^{\balpha}}{\partial \ba^{\balpha}}\Big|_{\ba=0} \int_\Omega \sigma(\ba\bx + \bb) \, d\mu(\bx)
\\
\label{eq:ProofS1.1_eqD} 
&= 
 \int_\Omega \frac{\partial^{\balpha} \sigma (\bx_0)}{\partial \bx^{\balpha}} \,\bx^{\balpha} \, d\mu(\bx)~.
\end{align}
Since $\sigma$ is non-degenerate, there exists $\bx_0\in \Omega$, such that $\partial^{\balpha} \sigma(\bx_0) / \partial \bx^{\balpha} \ne 0$. Thus, it follows that 
\begin{align}
\label{eq:ProofS1.1_eqE} 
\int_\Omega \bx^{\balpha} \, d\mu(\bx) = 0, \quad \forall \, \balpha \in \N^d_0~.
\end{align}
Above, the multi-index $\balpha$ was arbitrary. Multiplying the above identity \eqref{eq:ProofS1.1_eqE} by scalar coefficients $c_\alpha$ and summing a finite number of terms, it follows that $\int_\Omega p(\bx) \, d\mu(\bx) = 0$ for all multivariate polynomials $p(\bx)$. Because multivariate polynomials are dense in $C(\Omega)$, we conclude that $\int_\Omega f(\bx) \, d\mu(\bx) = 0$ for all continuous functions. This is only possible, if $\mu \equiv 0$. This shows that, if $\sigma$ is non-degenerate, PNNs of the form \eqref{eq:mPNN} are dense in $C(\Omega)$.

\indent To prove the other direction, we need to show that if $\sigma$ is degenerate, then the subspace $V$ consisting of all functions of the form \eqref{eq:mPNN} cannot be dense in $C(\Omega)$. By assumption on $\sigma$, there exists a multi-index $\balpha\in \N^d$, such that $D^{\balpha} \sigma := \partial^{\balpha}\sigma / \partial \bx^{\balpha} \equiv 0$. We can pick a smooth function $\phi$ with the following properties: $\phi(\bx) \in [0,1]$ for all $\bx\in \Omega$, $\phi(\bx) \equiv 0$ in a neighborhood of the boundary $\partial \Omega$, and $\phi(\bx) \equiv 1$ on an open ball $B\subset \Omega$. Then define a continuous linear functional
\begin{align}
\label{eq:ProofS1.1_eqF} 
\ell(f) := \int_{\Omega} f(\bx) D^{\balpha}\big(\bx^{\balpha} \phi(\bx)\big) \, d\bx~.
\end{align}
We first note that $\ell$ is non-trivial, since for any non-trivial continuous function $\psi \ge 0$ supported in $B\subset \Omega$, we have
\begin{align}
\ell(\psi) = \int_{B} \psi(\bx) D^{\balpha}\big(\bx^{\balpha} \underbrace{\phi(\bx)}_{\equiv 1}\big) \, d\bx
= \alpha! \int_{B} \psi(\bx) \, d\bx
> 0~.
\end{align}
Thus $\ell \ne 0$. On the other hand, integration by parts (taking into account that $\phi$ vanishes identically on $\partial \Omega$), yields
\begin{align}
\label{eq:ProofS1.1_eqH} 
\ell(v) 
&= \int_{\Omega} v(\bx) D^{\balpha} \big(\bx^{\balpha} \phi(\bx)\big) \, d\bx
\\
&= (-1)^{|\balpha|} \int_{\Omega} \big( D^{\balpha} v(\bx) \big) \bx^{\balpha} \phi(\bx) \, d\bx~.
\end{align}
By assumption on $\sigma$, the fact that $D^{\balpha} \sigma(\bx) \equiv 0$ implies that $D^{\balpha} v(\bx) \equiv 0$ for all $v\in V$, that is, functions $v(\bx)$ of the form \eqref{eq:mPNN}. Thus, it follows that $\ell(v) = 0$ for all $v\in V$, implying that $V$ cannot be dense by Proposition \ref{prop:dense}.
\end{proof}
\end{adjustwidth}
\noindent The above proof holds for sums over identical encoding functions $\sigma(\bx)$. However, in a real-world system we might have many (slightly) different encoding functions, yielding an expression of the form
\begin{align}
    \text{f}(\bx) &= \sum_{j=0}^{r} c_j\sigma_j(\ba_j \smallcirc\textbf{x}+\bb_j)
\label{eq:index}
\end{align}
with pairwise different encoding functions $\sigma_j$. In the following we will prove that universality is preserved if each of the encoding functions satisfies the degeneracy condition.
\begin{indentedtheorem}
\label{thm:universality_index}
Let $\Omega \subset \R^d$ be a compact domain and $\sigma_j \in C^\infty(\R^d)$ pairwise different analytical functions. Then PNNs of the form \eqref{eq:index} are universal in $C(\Omega; \R^m)$, if each $\sigma_j$ is non-degenerate in the following sense: no multi-index 
$\balpha = (\alpha_1,\dots, \alpha_d) \in \N_0^d$ exists such that $\partial^{\balpha} \sigma_j / \partial \bx^{\balpha} \equiv 0$.
\end{indentedtheorem}
\begin{adjustwidth}{3em}{0pt}
\begin{proof}[Proof of Theorem \ref{thm:universality_index}]
We will prove the above theorem by showing that functions of the form \eqref{eq:index} can uniformly approximate arbitrary polynomials in the limit $r\to \infty$ (with infinitely many pairwise different $\sigma_j$) by induction over the degree $\text{N}$.\\
\noindent $\textbf{N}=\mathbf{0:}$ We need to approximate a constant function. We note that for an arbitrary $\sigma_j(\bx)$ we can find $\bb_j$ such that $\sigma_j(\bb_j) \neq 0$ due to non-degeneracy. We can then set $\ba_j = 0$ and choose $c_j$ to approximate arbitrary constants as $\text{f}(\bx) = c_j \sigma_j(\bb_j)$.\\
$\textbf{Induction hypothesis:}$ We can approximate arbitrary polynomials of degree $\leq \text{N}$.\\
$\textbf{N}\mathbf{+1:}$ We note that the Taylor expansion of $\sigma_j(\bx)$ can be written as
\begin{align}
    \sigma_j(\ba_j\smallcirc\bx +\bb_j) = P_{k<N+1}(\bx) + \sum_{|\alpha| = N+1}p_\alpha(\bb_j) \ba_j^\alpha\bx^\alpha + P_{k>N+1}(\bx),
\label{eq:Taylor}
\end{align}

\noindent where we defined $\bx^\alpha = x_0^{\alpha_0}...x_d^{\alpha_d}$, $\ba_j^\alpha = a_{j0}^{\alpha_0}...a_{jd}^{\alpha_d}$, and $p_\alpha(\bb_j) \sim \frac{\partial^\alpha \sigma_j}{\partial\bx^\alpha}(\bb_j)$. Further, $P_{k<N+1}(\bx)$ describes a polynomial of degree smaller than $N+1$ and $P_{k>N+1}(\bx)$ is a power series with all constituents of degree larger than $N+1$. By the induction hypothesis, we can find a function $g(\bx)$ of the form \eqref{eq:index} such that the lower-order polynomials $P_{k<N+1}(\bx)$ can be uniformly approximated. This means that for each $\epsilon' >0$ we can find a $g(\bx)$ such that
\begin{align}
    f_j(\bx)&:= c_j\sigma_j(\ba_j\smallcirc \bx + \bb_j) - c_jg_j(\bx)\\
    &=c_j \sum_{|\alpha| = N+1}p_\alpha(\bb_j) \ba_j^\alpha\bx^\alpha + P_{k>N+1}(\bx) + O(c_j \epsilon'),
\label{eq:induction_approx}
\end{align}

\noindent By further setting $\ba_j \to \epsilon\ba_j$ and $c_j \to \frac{ c_j}{\epsilon^{N+1}}$ and for $g$ such that $\epsilon' \leq \epsilon^{N+2}$, we get
\begin{align}
    f_j(\bx) &= c_j \sum_{|\alpha| = N+1}p_\alpha(\bb_j) \ba_j^\alpha\bx^\alpha + O(c_j \epsilon),
\label{eq:polynomial}
\end{align}

\noindent where $\epsilon >0$ can be made arbitrarily small. To complete the proof, we now need to show that a sum over expressions \eqref{eq:polynomial} can uniformly approximate the individual monomials $m_\alpha \bx^\alpha$. For this we set $a_{jk} = \delta^{(N+1)^{k}}$, such that $\ba_j^\alpha = \delta^{\alpha_0}\delta^{(N+1)\alpha_1}...\delta^{(N+1)^{d}\alpha_d}$, and we get
\begin{align}
    f_j(\bx) &= c_j \sum_{|\alpha| = N+1}p_\alpha(\bb_j) \delta^{q(\alpha)} \bx^\alpha + O(c_j \epsilon),
\label{eq:scaling_sum}
\end{align}

\noindent with $q(\alpha) = \sum_{k=0}^d (N+1)^k\alpha_k$ and $\alpha_0 +... \alpha_d = N+1$. We note that the exponents $q(\alpha)$ are distinct for distinct $\alpha$. Let $q_0:= q(\alpha^{(0)}) = N+1$ be the lowest order in $\delta$ present in \eqref{eq:scaling_sum}(the monomial $x_0^{N+1}$). We set $c_j\to \frac{c_j}{\delta^{q_0}}$ and pick $\epsilon \leq \delta^{q_0+1}$ to get
\begin{align}
    f_j(\bx) &= c_j p_{\alpha^{(0)}}(\bb_j)x_0^{N+1} + O(\delta).
\label{eq:monomial}
\end{align}
We note that $\bb_j$ can be chosen such that $p_\alpha(\bb_j)\sim \frac{\partial^\alpha \sigma_j}{\partial\bx^\alpha}(\bb_j)\neq0$ for the relevant coefficient due to the non-degeneracy. By drawing the limit $\delta\to0$ it follows that we can uniformly approximate arbitrary monomials $mx_0^{N+1}$. By subtracting such a monomial, it follows (for a fresh index $j$) that we can uniformly approximate expression \eqref{eq:scaling_sum} without the term proportional to $x_0^{N+1}$. We can then repeat the above argument to uniformly approximate the next lowest order (the monomial $x_0^{N}x_1$). By iteratively repeating this procedure it follows that we can uniformly approximate all monomials of degree $N+1$. This can be combined with the induction hypothesis to show that the expression \eqref{eq:index} can uniformly approximate arbitrary polynomials of degree $N+1$. This concludes the proof by induction, and it follows that the sum \eqref{eq:index} can uniformly approximate arbitrary multivariate polynomials. Since polynomials are dense in $C(\Omega)$, the PNNs of the form \eqref{eq:index} are universal.
\end{proof}
\end{adjustwidth}
%
%
\section{Free-space optical-system theory}
In the following we will derive a mathematical form of the encoding block in the case of general mirror reflectivities. \\
\indent For the general case, we assume that we have a partially reflective mirror with a complex transmission coefficient $\text{t}_\text{m}$ and reflection coefficient $\text{r}_\text{m}$. Furthermore, we assume the propagation through the pseudo-cavity is described by a matrix $\textbf{U}(\bx)$. For an incoming wave $\textbf{E}^{\text{in}}$, the output beam is then
\begin{align}
    \textbf{E}^{\text{out}} &= \text{r}_\text{m}\textbf{E}^{\text{in}} + \text{t}_\text{m}\textbf{U}(\bx)\text{t}_\text{m} \textbf{E}^{\text{in}} + \text{t}_\text{m}\textbf{U}(\bx)\text{r}_\text{m}\textbf{U}(\bx)\text{t}_\text{m}\textbf{E}^{\text{in}} + ... \\
    &= \big(\text{r}_\text{m}\I+ \text{t}_\text{m}^2\textbf{U}(\bx) + \text{r}_\text{m} \text{t}_\text{m}^2\textbf{U}(\bx)^2 +...\big)\textbf{E}^{\text{in}} \\
    &= \left(\text{r}_\text{m}\I + \text{t}_\text{m}^2 \sum_{n=1}^{\infty} \text{r}_\text{m}^{n-1} \textbf{U}(\bx)^n\right) \textbf{E}^{\text{in}} \\
    &\mkern-8mu\overset{\text{r}_\text{m} \neq 0}{=} \left(\text{r}_\text{m}\I + \frac{\text{t}_\text{m}^2}{\text{r}_\text{m}} \sum_{n=1}^{\infty} \text{r}_\text{m}^{n} \textbf{U}(\bx)^n\right) \textbf{E}^{\text{in}} \\
    &= \left( \left(\text{r}_\text{m} - \frac{\text{t}_\text{m}^2}{\text{r}_\text{m}}\right) \I + \frac{\text{t}_\text{m}^2}{\text{r}_\text{m}} \big(\I - \text{r}_\text{m}\textbf{U}(\bx)\big)^{-1} \right)\textbf{E}^{\text{in}}~,
\end{align}
\hspace{2.0em}
We note that in the proof and the main text we treat the case $\text{r}_\text{m}=\text{t}_\text{m}$ for which the above expression reduces to 
\begin{align}
\label{eq:encoding_simplified}
    \textbf{E}^{\text{out}} &= \text{r}_\text{m} \big(\I - \text{r}_\text{m}\textbf{U}(\bx)\big)^{-1}\textbf{E}^{\text{in}}~.
\end{align}
However, we note that the following proof easily extends to the case $\text{t}_\text{m}\neq\text{r}_\text{m}$, as the additional background field added by the first term is constant in $\bx$ and does hence not affect the universality criterion. \\
\indent Also, in the main text and the following sections we assume $\textbf{U}(\bx) = \textbf{T}(\bx)\textbf{S}\textbf{T}(\bx)$, where $\textbf{T}(\bx)$ is a diagonal phase or amplitude modulation and $\textbf{S}$ is a symmetric, unitary matrix. This treatment assumes monochromatic, coherent, paraxial optics that is both reciprocal and lossless. Recall that, for simplicity, \textbf{S} includes propagation from the SLM to the scattering surface and back to the SLM.
\vspace{-1em}
\begin{indentedremark}
    We note that for a single, reflective scattering surface \textup{$\textbf{S}_1$}, \textup{$\textbf{S}$} has the form \textup{$\textbf{S} = \textbf{H}\textbf{S}_1\textbf{H}$}, where \textup{$\textbf{H}$} is a Toeplitz matrix describing free-space propagation and \textup{$\textbf{S}_1$} is a diagonal matrix describing the phase profile imprinted on the beam by the scattering surface. We note that more general cases of \textup{$\textbf{S}$} could be considered, for example by using a stack of (transmissive) surfaces instead of only one. If necessary, propagation losses and absorption could also be included in \textup{$\textbf{H}$} and \textup{$\textbf{S}_1$}, respectively.
\end{indentedremark}
\vspace{-1em}
\begin{indentedremark}
    For clarity we note that for a lossless, 50\% reflective mirror we would have $|\text{t}_\text{m}| = |\text{r}_\text{m}|$ but generally $\text{t}_\text{m} \neq \text{r}_\text{m}$ as the reflection coefficient would carry a $\pi/2$ phase shift. The case we treat (no phase shift, equal transmission and reflection) is physically realized when using partially absorbing metallic mirrors, that is, $\text{t}_\text{m} = \text{r}_\text{m}$ but $|\text{r}_\text{m}|^2 + |\text{t}_\text{m}|^2<1$. However, as noted above, the proof extends to the general case.
\end{indentedremark}
\section{Universality of proposed free-space optical system}
\label{app:free_space}
The main goal of this section is to use the universality theorem (Theorem \ref{thm:universality}) to prove that the free-space optical system proposed in the main text is universal at large scale. For this, recall that the input--output mapping for the proposed system behind each lens is mathematically given by the expression:
\begin{align}
\label{eq:ProofS1.1_eqG} 
\mathbf{E}^\text{out}_k = \mathcal{F}\left\{ \bSigma(\bx;\bS) \mathbf{E}^\text{in}_k\right\}~,
\end{align}
where
\begin{align}
\bSigma(\bx;\bS) 
&:= \text{r}_\text{m} \big( \I - \text{r}_\text{m}\bT(\bx)\bS\bT(\bx)\big)^{-1}~.
\end{align}
\noindent For the proof, we will restrict ourselves to a single lens and one output variable. The extension to multiple variables follows directly by assuming an additional lens for each output variable (see Remark \ref{rmk:multilens} below). In the following, we will drop the subscript $k$ for simplicity, that is, $\mathbf{E}^\text{in}_k \to \mathbf{E}^\text{in}$ and $\mathbf{E}^\text{out}_k \to \mathbf{E}^\text{out}$
\subsection{Non-degeneracy of the nonlinear encoding function, \texorpdfstring{$\sigma$}{sigma}}
\noindent For the moment, we consider a diagonal embedding of the inputs in matrix $\bT$, that is $\bT(\bx) = \mathrm{diag}(x_1,\dots, x_r)$. Later $\bx$ will be adjusted using the form $\ba_j\smallcirc\bx + \bb_j$. We further note that this case corresponds to amplitude modulation by the SLM. It will be extended to phase modulation in Section \ref{sec:phase_modulation}.

To begin our analysis, we choose $\mathbf{E}^\text{in} = [1, 1, \dots, 1]
^{\mathrm{T}}$. On the output side, we assume that only the zeroth Fourier component of the output is measured. Under the assumption that there is only one input copy in the system, this results in nonlinear encoding functions of the form (shown in \ref{thm:universality_PNN})
\vspace{-1em}
\begin{align}
\label{eq:act}
\sigma(\bx) 
= \sigma(\bx;\bS)
:= \mathbf{e}^{\mathrm{T}} \cdot \bSigma(\bx;\bS) \cdot \mathbf{e}~, \quad \mathbf{e} = [1,\dots, 1]^{\mathrm{T}}~.
\end{align}
Our first aim is to show that $\sigma(\bx;\bS)$ is non-degenerate (in the sense of Theorem \ref{thm:universality}) for generic $\bS$. In subsequent sections we will discuss the physical realization of a universal architecture based on this non-degeneracy.\\
\indent \textbf{Restrictions on $\bS$.}
Consistent with physical constraints, we will only consider unitary and symmetric $\bS$, that is, we assume that $\bS$ satisfies $\bS^\dagger \bS = \I$ and $\bS^{\mathrm{T}} = \bS$. Any such $\bS$ can be represented in the form $\bS = \exp(i\bB)$, where $\bB$ is a real--symmetric matrix. Conversely, for any real--symmetric $\bB$, we have that $\bS = \exp(i\bB)$ is unitary and symmetric. 
\\
\indent \textbf{Non-degeneracy of the nonlinear encoding function.}
We then have the following result:
\vspace{-1.5em}
\begin{indentedtheorem}
\label{thm:nondegenerate}
The nonlinear encoding function $\sigma(\bx;\bS)$ given by expression \eqref{eq:act} fulfills the universality criterion for almost all $\bS$; more precisely, the set of real--symmetric $\bB$, such that $\bS = \exp(i\bB)$ gives rise to a non-degenerate $\sigma(\bx;\bS)$, has full Lebesgue measure\cite{Rudin1987}.
\end{indentedtheorem}
\subsection{Proof of Theorem \ref{thm:nondegenerate}}
\noindent We will show that for almost all real--symmetric $\bB$ that satisfy $\bS = \exp(i\bB)$, the function $\sigma(\bx;\bS)$ is non-degenerate (specifically, in a neighborhood of $\bx=0$).
\begin{indentedlemma}
\vspace{-1em}
\label{lem:generic}
Let $r,d\in \N$ be given. Then for Lebesgue-almost all real--symmetric $\bB \in \R^{d\times d}$, we have that $\bS = \exp(i\bB)$ satisfies,
\vspace{+0.4em}
\begin{align}
\label{eq:lemmaS2.1}    
\frac{\partial^{2rd}}{\partial x_1^{2r} \dots \partial x_d^{2r}}
\Big|_{\bx=0}
\sigma(\bx;\bS)
\ne 0~.
\end{align}
\end{indentedlemma}
\begin{indentedremark}
\vspace{-1.0em}
Lemma \ref{lem:generic} is phrased for a single, fixed $r$. But since a countable union of sets of measure zero has itself measure zero, we can conclude that for almost all $\bB \in \mathrm{Sym}(d)$, we have that $\bS = \exp(i\bB)$ satisfies,
\begin{align}
\label{eq:remarkS2.1}    
\frac{\partial^{2rd}}{\partial x_1^{2r} \dots \partial x_d^{2r}}
\Big|_{\bx=0}
\sigma(\bx;\bS)
\ne 0~,
\quad
\text{for all $r\in \N$~.}
\end{align}
Hence, Lemma \ref{lem:generic} shows that $\sigma(\bx;\bS)$ is generically non-degenerate for unitary symmetric $\bS$.
\end{indentedremark}
\begin{indentedremark}
\vspace{-1.3em}
\label{rem:generic}
Non-degeneracy for arbitrary $\balpha$ is immediate from Lemma \ref{lem:generic}, since $\partial^{\balpha} \sigma(\bx;\bS) / \partial \bx^{\balpha} \equiv 0$ would also imply that 
\vspace{-0.0em}
\begin{align}
\label{eq:remarkS2.2}   
\frac{\partial^{2rd}}{\partial x_1^{2r} \dots \partial x_d^{2r}}
\sigma(\bx;\bS) \equiv 0~,
\end{align}
for all sufficiently large $r$. This is clearly ruled out by Lemma \ref{lem:generic}. Thus, Lemma \ref{lem:generic} immediately implies Theorem \ref{thm:nondegenerate}.
\end{indentedremark}
\vspace{+0.5em}
Our next aim is to prove Lemma \ref{lem:generic}. To this end, we first note that 
\begin{align}
\label{eq:expansion}
\mathbf{e}^{\mathrm{T}} \cdot \left[
\bT(\bx)\bS\bT(\bx)
\right]^k
\cdot 
\mathbf{e} = \sum_{j_0,j_1,\dots, j_k} S_{j_0 j_1} \dots S_{j_{k-1} j_k} 
\; x_{j_0} x_{j_1}^2 \dots x_{j_{k-1}}^2 x_{j_k}~,
\end{align}
where the sum is over all $j_0,\dots, j_k \in \{1,\dots, d\}$. In particular, equation \eqref{eq:expansion} defines a $2k$-homogeneous polynomial. From this and the Neumann series expansion of 
\begin{align}
\label{eq:Neumann_exp}
\sigma(\bx;\bS) = \text{r}_\text{m} \sum_{k=0}^\infty \e^{\mathrm{T}} \cdot \big(\text{r}_\text{m}\bT(\bx)\bS\bT(\bx)\big)^k \cdot \e~,
\end{align}
it follows that
\vspace{+0.5em}
\begin{align}
\frac{\partial^{2rd}}{\partial x_1^{2r} \dots \partial x_d^{2r}}
\Big|_{\bx=0}
\sigma(\bx;\bS)\;=\;
\frac{\partial^{2rd}}{\partial x_1^{2r} \dots \partial x_d^{2r}}
\Big|_{\bx=0}
\,\text{r}_\text{m} \,\e^{\mathrm{T}} \cdot \big(\text{r}_\text{m}\bT(\bx)\bS\bT(\bx)\big)^{2rd} \cdot \e~.
\end{align}
\noindent When taking derivatives of even order (in each coordinate direction) of the expansion \eqref{eq:expansion}, then only terms with $j_0 = j_k$ will yield a non-zero contribution. Because of this, it follows that
\begin{align}
&\frac{\partial^{2rd}}{\partial x_1^{2r} \dots \partial x_d^{2r}}
\Big|_{\bx=0}
\sigma(\bx;\bS)
\nonumber \\
&= 
\text{r}_\text{m}^{2rd+1}
\frac{\partial^{2rd}}{\partial x_1^{2r} \dots \partial x_d^{2r}}
\Big\{
\sum_{j_1,\dots, j_{rd}}
S_{j_1 j_2} \dots S_{j_{rd-1} j_{rd}} S_{j_{rd},j_1}
\; 
x_{j_1}^2 \dots x_{j_{rd-1}}^2 x_{j_{rd}}^2
\Big\}~.
\end{align}
\noindent Let us now define $\mathcal{P}_r^d$ as a set of sequences of length $rd$, consisting of all sequences \linebreak 
$\mathbf{j}=(j_1,\dots, j_{rd}) \in \{1,\dots, d\}^{rd}$ such that $\mathbf{j}$ contains \emph{exactly} $r$ copies of each index in $\{1,\dots, d\}$. Or to be more explicit, $\mathcal{P}_r^d$ contains
\begin{align}
(\underbrace{1,\dots, 1}_{r \text{ times}}, \underbrace{2,\dots, 2}_{r \text{ times}}, \dots, \underbrace{d,\dots, d}_{r \text{ times}}) \in \mathcal{P}_r^d~,
\end{align}
and all possible permutations of this sequence. The motivation for introducing $\mathcal{P}_r^d$ is that we have
\begin{align}
\frac{\partial^{2rd}}{\partial x_1^{2r} \dots \partial x_d^{2r}}\Big|_{x=0} x_{j_1}^2 \dots x_{j_{rd}}^2
&=
\frac{\partial^{2rd}}{\partial x_1^{2r} \dots \partial x_d^{2r}}\Big|_{x=0} x_{1}^{2r} \dots x_{d}^{2r}
\\
&\ne 0~, \qquad \forall~ \mathbf{j} \in \mathcal{P}_r^d~,
\end{align}
whereas,
\begin{align}
\frac{\partial^{2rd}}{\partial x_1^{2r} \dots \partial x_d^{2r}}\Big|_{x=0} x_{j_1}^2 \dots x_{j_{rd}}^2
= 0~, \quad \forall ~\mathbf{j} \notin \mathcal{P}_r^d~.
\end{align}
From the above expansion of $\sigma(\bx;\bS)$, it thus now follows that
\begin{align}
&\frac{\partial^{2rd}}{\partial x_1^{2r} \dots \partial x_d^{2r}}
\Big|_{\bx=0}
\sigma(\bx;\bS)
\nonumber
\\
&= 
\text{r}_\text{m}^{2rd+1}
\frac{\partial^{2rd}}{\partial x_1^{2r} \dots \partial x_d^{2r}}
\Big\{
\sum_{{\mathbf{j}\in \mathcal{P}_r^d}}
S_{j_1 j_2} \dots S_{j_{rd-1} j_{rd}} S_{j_{rd},j_1}
\; 
x_{j_1}^2 \dots x_{j_{rd-1}}^2 x_{j_{rd}}^2
\Big\}
\\
&= 
\text{r}_\text{m}^{2rd+1}
\frac{\partial^{2rd}}{\partial x_1^{2r} \dots \partial x_d^{2r}}
\Big\{
\sum_{\mathbf{j}\in \mathcal{P}_r^d}
S_{j_1 j_2} \dots S_{j_{rd-1} j_{rd}} S_{j_{rd},j_1}
\; 
x_{1}^{2r} \dots x_{d}^{2r}
\Big\}
\\
&= 
\text{r}_\text{m}^{2rd+1}[(2r)!]^d
\sum_{(j_1,\dots, j_{rd}) \in \mathcal{P}_r^d}
S_{j_1 j_2} \dots S_{j_{rd-1} j_{rd}} S_{j_{rd},j_1}~.
\end{align}
Thus, to prove that 
\begin{align}
\frac{\partial^{2rd}}{\partial x_1^{2r} \dots \partial x_d^{2r}}
\Big|_{\bx=0}
\sigma(\bx;\bS)
\ne 0~,
\end{align}
for almost all $\bS = \exp(i\bB)$, it suffices to show that the following polynomial in the components of $\bS$ is non-vanishing for almost all real-symmetric $\bB \in \R^{d\times d}$:
\begin{align}
\label{eq:nondeg0}
p_{r,d}(\bS) 
:= 
\sum_{(j_1,\dots, j_{rd}) \in \mathcal{P}_r^d}
S_{j_1 j_2} \dots S_{j_{rd-1} j_{rd}} S_{j_{rd},j_1}~.
\end{align}
To prove this, we first consider $r=1$ and note that any sequence $(j_1,\dots, j_d) \in \mathcal{P}_1^d$ is a permutation of $(1,2,\dots, d)$ and thus has no repeated entries. Hence, the sum 
\begin{align}
p_{1,d}(\bS) 
= 
\sum_{(j_1,\dots, j_{d}) \in \mathcal{P}_1^d}
S_{j_1 j_2} \dots S_{j_{d-1} j_{d}} S_{j_{d},j_1}~,
\end{align}
only involves off-diagonal components $S_{ij}$ with $i\ne j$. \\
\indent Given $\bB\in \R^{d\times d}$, and $\tau \in \R$, we now consider $\bS(\tau) := \exp(i\tau \bB)$. We note that $\bS(0) = \I$. In particular, all off-diagonal components of $\bS(0)$ vanish, implying that $p_{1,d}\big(\bS(0)\big) = 0$. In fact, upon Taylor expanding $p_d\big(\bS(\tau)\big)$ around $\tau=0$, we find
\begin{align}
p_{1,d}\big(\bS(\tau)\big) = i^d \tau^d p_{1,d}(\bB) + \text{(higher-order terms in $\tau$)}~.
\end{align}
This expression shows that for sufficiently small $\tau > 0$, any real-symmetric $\bB$ for which $p_{1,d}(\bB) \ne 0$ leads to a $p_{1,d}\big(\bS(\tau)\big) \ne 0$. However, since $p_{1,d}(\bB)$ is a non-trivial polynomial in the components of $\bB$, the following result already implies that $p_{1,d}(\bB) \ne 0$ for (Lebesgue-) almost all real-symmetric $\bB \in \R^{d\times d}$, upon identifying the space of real-symmetric matrices $\bB\in \R^{d\times d}$ with the Euclidean space $\bm{\xi} \in \R^{d(d+1)/2}$:  
\vspace{-1em}
\begin{indentedlemma}
If $p: \R^n \to \R$ is a multivariate polynomial, then the set $\{\bm{\xi} \in \R^n \, :\, p(\bm{\xi}) = 0\}$ has Lebesgue measure zero, unless $p(\bm{\xi}) \equiv 0$.
\end{indentedlemma}
\vspace{+1em}

In particular, it follows that for almost all $\bB$, we have that the composition \linebreak $\tau \mapsto p_{1,d}\big(\bS(\tau)\big) = p_{1,d}\big(\exp(i\tau \bB)\big)$ is a non-trivial function. In addition, it is a holomorphic (even entire) function. Thus for any $\bB$ for which $p_{1,d}\big(\bS(\tau)\big) \not \equiv 0$, this holomorphy implies that the set of zeros in $\tau$, that is,
\begin{align}
\{ \tau >0 \, :\, p_{1,d}\big(\bS(\tau)\big) = p_{1,d}\big(\exp(i\tau \bB)\big) = 0 \}~,
\end{align}
consists of isolated points, as the zeros of holomorphic functions cannot have accumulation points inside their domain of definition. Thus, denoting by $\mathrm{Sym}(d) \simeq \R^{d(d+1)/2}$ the set of real-symmetric matrices, we arrive at the following observation: For almost every $\bB\in \mathrm{Sym}(d)$ and for almost every $\tau > 0$, we have that $p_{1,d}\big(\exp(i\tau\bB)\big) \ne 0$. From this, it follows that
\begin{align}
\{ \bB \in \mathrm{Sym}(d) \, : \, p_{1,d}\big(\exp(i\bB)\big) = 0 \}~,
\end{align}
must have zero Lebesgue-measure. This shows the claimed non-degeneracy of equation \eqref{eq:nondeg0} for $r = 1$.\\
\indent In the more general case, we note that we can write
\begin{align}
\label{eq:prd-expansion}
p_{r,d}\big(\exp(i\tau \bB)\big)
= r \, i^d \tau^d p_{1,d}(\bB) + O(\tau^{d+1})~.
\end{align}
To see this, we note that for any index-sequence $(j_1,\dots, j_{rd}) \in \mathcal{P}_r^d$, a consecutive pair of indices $(j_{k},j_{k+1})$ introduces a factor $S_{j_k,j_{k+1}} = \delta_{j_k,j_{k+1}} + i\tau B_{j_k,j_{k+1}} + O(\tau^2)$ in equation \eqref{eq:nondeg0}. The lowest order term $O(\tau^d)$ is thus obtained by minimizing the number of transitions with $j_k \ne j_{k+1}$, which corresponds to permutations of $(1,\dots,1,2,\dots,2,\dots, d,\dots, d)$ for which the blocks of $1$'s, $2$'s, etc.\,are kept contiguous. Up to $r$ cyclic left-shifts, one can uniquely identify such a contiguous sequence with a sequence in $\mathcal{P}^d_1$. Indeed, this follows by considering the mapping $(j_1, \dots, j_{rd}) \mapsto (j_1,j_{r+1},\dots, j_{(r-1)d+1})$. This then leads to equation \eqref{eq:prd-expansion}.

We have already shown above that $p_{1,d}(\bB) \ne 0$ for almost all $\bB \in \mathrm{Sym}(d)$. Thus, we can argue along similar lines also for $r>1$: The expansion \eqref{eq:prd-expansion} shows that, for almost all real-symmetric $\bB$, we have $p_{r,d}\big(\exp(i\tau \bB)\big) \ne 0$ for small $\tau > 0$. This again implies that $\tau \mapsto p_{r,d}\big(\exp(i\tau \bB)\big)$ is a non-trivial holomorphic function, and hence all zeroes must be isolated from each other. In particular, the zeros in $\tau$ form a set of Lebesgue-measure zero. This is again sufficient to deduce that in fact the set
\begin{align}
\{ \bB \in \mathrm{Sym}(d) \, : \, p_{r,d}\big(\exp(i\bB)\big) = 0 \}~,
\end{align}
must have zero Lebesgue-measure, implying the claimed non-degeneracy in equation \eqref{eq:nondeg0} also for $r > 1$. \\ 
\indent This completes our proof of non-degeneracy, Lemma \ref{lem:generic}. As pointed out in Remark \ref{rem:generic}, this lemma immediately implies Theorem \ref{thm:nondegenerate}.

\begin{indentedremark}
    We note that the above proof holds for unitary, symmetric $\textbf{S}$. In an experiment, propagation and absorption losses could break the unitarity. However, we expect the argument could be extended to the account for the more general case. For example, if the system matrix can be written as $\tilde{\textbf{S}} = \gamma\textbf{S}$, where $0<\gamma<1$ is an attenuation-factor accounting for losses, the extension follows directly. 
\end{indentedremark}
\subsection{Extension to phase-modulation encoding}
\label{sec:phase_modulation}
\noindent In definition \eqref{eq:act}, it is assumed that the input encoding $\bT(\bx) = \mathrm{diag}(x_1, \dots, x_r)$ is performed by amplitude modulation. If we instead consider input encoding via phase modulation, then this corresponds to the choice $\widetilde{\bT}(\bphi) = \mathrm{diag}(e^{i \phi_1}, \dots, e^{i \phi_r})$, equivalent to the parametrization $\bx = e^{i\bphi}$. In this case, we are interested in the non-degeneracy of the (complex-valued) nonlinear encoding function
\begin{align}
\label{eq:phaseact}
\sigma(\bphi; \bS) = \sigma(e^{i\bphi}; \bS)~,
\end{align}
written in terms of phase-angles $\bphi = (\phi_1, \dots, \phi_r)$.
The main difficulty when moving from $x_j$ to $\phi_j$ is that the derivation of the universality for $\bT(\bx) = \mathrm{diag}(x_1,\dots, x_r)$ relied on taking derivatives in the $x_j$ and specifically setting $x_j = 0$ to show non-degeneracy. With the parametrization $x_j = e^{i\phi_j}$, we can no longer evaluate the derivatives at $x_j=0$. Nevertheless, we can prove the following non-degeneracy result:
\vspace{-1.0em}
\begin{indentedtheorem}
\label{thm:nondegenerate-phase}
If $\sigma(\bphi;\bS)$ denotes the multivariate encoding function \eqref{eq:phaseact}, obtained by employing phase-encoding $\bx = e^{i\bphi}$ in equation \eqref{eq:act}, then $\sigma(\bphi;\bS)$ is non-degenerate for almost all $\bS$; more precisely, for almost all real-symmetric $\bB$, the matrix $\bS = \exp(i\bB)$ satisfies the non-degeneracy condition,
\begin{align}
\frac{\partial^{\balpha} \sigma(\bphi;\bS)}{\partial \bphi^{\balpha}} \not \equiv 0~,
\qquad \forall \, \balpha \in \N_0^d~.
\end{align}
\end{indentedtheorem}
\begin{adjustwidth}{3em}{0pt} 
\begin{proof}[Proof of Theorem \ref{thm:nondegenerate-phase}]
For the same reason as in the proof of Theorem \ref{thm:nondegenerate} based on Lemma \ref{lem:generic}, it suffices to show that for almost all $\bS$, we have
\begin{align}
\frac{\partial^{2rd} \sigma(\bphi;\bS)}{\partial \phi_1^{2r} \dots \partial \phi_d^{2r}} \not \equiv 0~,
\end{align}
for all $r\in \N$. We will base our proof of non-degeneracy on the previous results for the amplitude encoding, $\bT(\bx) = \mathrm{diag}(x_1,\dots, x_r)$.

From Lemma \ref{lem:generic}, we already know that for almost all unitary and symmetric $\bS$, the function $\bx \mapsto \sigma(\bx;\bS)$ is non-degenerate at $\bx=0$, more precisely,
\begin{align}
\label{eq:nondeg-at-0}
\frac{\partial^{2rd}}{\partial x_1^{2r} \dots \partial x_d^{2r}}
\Big|_{\bx=0}
\sigma(\bx;\bS)
\ne 0~,
\quad
\text{for all $r\in \N$~.}
\end{align}
In the following, we will keep such $\bS$ fixed and show that for $\bS$ satisfying this non-degeneracy condition, the phase-encoded activation $\sigma(\bphi; \bS)$ is also non-degenerate as a function of $\bphi$. To simplify notation, we will suppress the additional dependency on $\bS$ in the following.
By the definition of $\sigma(\bx) := \sigma(\bx;\bS)$, it also follows that the Neumann series defining $\sigma(\bx)$ converges for any $\bx$ with components $|x_j| < |\frac{1}{\text{r}_\text{m}}|$. In fact, we have convergence of the series even when replacing $x_j$ by complex-valued $z_j$ with $|z_j|<|\frac{1}{\text{r}_\text{m}}|$, and hence $\sigma(\cdot)$ defines a non-trivial holomorphic function,
\begin{align}
\sigma: 
\underbrace{
\mathbb{D}\times \dots \times \mathbb{D}
}_{
\text{$d$-fold product}
}
\to \mathbb{C},
\quad (z_1, \dots, z_d) \mapsto \sigma(\bz)~,
\end{align}
where $\mathbb{D} = \{\zeta\in \mathbb{C}\,|\, |\zeta|<|\frac{1}{\text{r}_\text{m}}|\}$ is the disk of radius $|\frac{1}{\text{r}_\text{m}}| > 1$ in the complex plane $\mathbb{C}$.
This implies that we have a convergent series expansion of the form 
\begin{align}
\sigma(\bz)
=
\sum_{\balpha} c_{\balpha} \bz^{\balpha}~,
\end{align}
where $\bz^{\balpha} = z_1^{\alpha_1}\dots z_d^{\alpha_d}$, for $|z_j|<2$. By the known non-degeneracy in equation \eqref{eq:nondeg-at-0} of $\sigma$ at $0$, then $c_\alpha \ne 0$ for all $\balpha = (2r, 2r, \dots, 2r)$, since in this case
\begin{align}
\label{eq:nonvanish}
c_{\balpha} = \frac{1}{\balpha!} \frac{\partial^{\balpha}\sigma(\bx)}{\partial \bx^{\balpha}}\Big|_{\bx=0} \ne 0~,
\quad
\forall \, \balpha = (2r, \dots, 2r)~.
\end{align}
Substitution of $\bz = e^{i\bphi}$ into the above series expansion, yields
\begin{align}
\sigma(\bphi)
=
\sum_{\balpha} c_{\balpha} e^{i(\balpha\cdot\bphi)}~,
\end{align}
where the sum is over all $\balpha = (\alpha_1,\dots, \alpha_d)$ with $\alpha_1,\dots, \alpha_d \in \N_0$.
Fixing a multi-index $\bn = (2r, \dots, 2r)$, and taking partial derivatives, we arrive at
\begin{align}
\frac{\partial^{2rd}}{\partial \phi_1^{2r} \dots \partial \phi_d^{2r}} \sigma(\bphi)
&=
\frac{\partial^{2rd}}{\partial \phi_1^{2r} \dots \partial \phi_d^{2r}} 
\sum_{\balpha} c_{\balpha} e^{i(\balpha\cdot\bphi)}
\\
&= 
\sum_{\balpha} c_{\balpha} i^{2rd} \left( \prod_{j=1}^d \alpha_j^{2r} \right) e^{i(\balpha\cdot\bphi)}~.
\end{align}
\indent To show that $\frac{\partial^{2rd}}{\partial \phi_1^{2r} \dots \partial \phi_d^{2r}} \sigma(\bphi)$ does not vanish identically, we now integrate it against $e^{-i(\bn\cdot \bphi)}$ over the $d$-dimensional torus $\mathbb{T}^d$. This yields
\begin{align}
\int_{\mathbb{T}^d} 
e^{-i(\bn\cdot \bphi)} \frac{\partial^{2rd} \sigma(\bphi)}{\partial \phi_1^{2r} \dots \partial \phi_d^{2r}} 
\, d\bphi
&= 
\sum_\alpha c_\alpha i^{2rd} \left( \prod_{j=1}^d \alpha_j^{2r} \right) 
\underbrace{
\int_{\mathbb{T}^d} 
e^{-i(\bn\cdot \bphi)}e^{i(\balpha\cdot\bphi)}
\, d\phi
}_{
= (2\pi)^d \delta_{2r,\alpha_1} \dots \delta_{2r,\alpha_d}
}
\\
&= 
\label{eq:two_pi}
(2\pi)^d  i^{2rd}(2r)^{2rd} c_{\bn}~.
\end{align}
Since $c_{\bn} \ne 0$ for $\bn=(2r,\dots, 2r)$ by equation \eqref{eq:nonvanish}, the expression in equation \eqref{eq:two_pi} is not equal to zero, and hence, we have shown that 
\begin{align}
\int_{\mathbb{T}^d} 
e^{-i(\bn\cdot \bphi)} \frac{\partial^{2rd} \sigma(\bphi)}{\partial \phi_1^{2r} \dots \partial \phi_d^{2r}} 
\, d\bphi
\ne 0~.
\end{align}
Clearly, this is only possible if $ \frac{\partial^{2rd}}{\partial \phi_1^{2r} \dots \partial \phi_d^{2r}} \sigma(\bphi) = \frac{\partial^{2rd}}{\partial \phi_1^{2r} \dots \partial \phi_d^{2r}} \sigma(\bphi;\bS)$ doesn't vanish identically. This concludes our proof of the non-degeneracy Theorem \ref{thm:nondegenerate-phase}, for the physical activation function with  phase encoding, $\bx = e^{i\bphi}$.
\end{proof}
\end{adjustwidth}
\subsection{Linear recombination}\label{sec:linear_recombination}
\noindent In the above sections, we proved that encoding functions of the form
\begin{align}
\label{eq:act_v2}
\sigma_j(\bx)
:= \text{r}_\text{m}~\mathbf{e}_{j}^{\mathrm{T}} \cdot \big( \I - \text{r}_\text{m}\bT(\bx)\bS_j\bT(\bx) \big)^{-1} \cdot \mathbf{e}_{j}
\end{align}
satisfy the universality criterion for $\mathbf{e}_{j} = [1,\dots, 1]^{\mathrm{T}}$ and almost all $\bS_j$. We have introduced a subscript $j$ to distinguish encoding functions corresponding to different input copies. To conclude universality of the proposed PNN, we must now show that we can encode the input multiple times in the system and then linearly recombine all components as required by the universality theorem. \\
\indent Specifically, behind each lens we must obtain a one dimensional output of the form
\begin{align}
\label{eq:fdecoupled}
    f(\bx) = \sum_j c_j \sigma_j(\ba_j\bx + \bb_j)~,
\end{align}
for arbitrary values of $\ba_j$, $\bb_j$, and $c_j$. For this, we first recall that the beam profile in the detector plane behind each lens is given by the expression
\begin{align}
    \textbf{E}^{\text{out}} = \mathcal{F}\left\{\text{r}_\text{m} \big( \I - \text{r}_\text{m} \textbf{T}(\bx) \textbf{S} \textbf{T}(\bx) \big)^{-1} \textbf{E}^\text{in}\right\}.
\label{eq:reminder}
\end{align}
We note that the above Fourier transform is applied in two dimensions (2D) to respect the spatial structure of the beam. \\
\indent For the proof, we will assume that the input replications are spatially separated on the SLM, such that the system effectively decouples into separate sub-blocks, each of which contains one `copy' of the input. For $n$ such copies, equation \eqref{eq:reminder} takes the form
\begin{align}
\label{eq:decoupledPNN}
     \mathbf{E}^{\text{out}}(\bx) &= \mathcal{F}\left\{\begin{pmatrix}
    \text{r}_\text{m} \left( \I - \text{r}_\text{m}\textbf{T}(\bx) \textbf{S}_1 \textbf{T}(\bx) \right)^{-1} \text{E}^{(1)} \\
    \vdots \\
    \text{r}_\text{m} \left( \I - \text{r}_\text{m}\textbf{T}(\bx) \textbf{S}_n \textbf{T}(\bx) \right)^{-1} \text{E}^{(n)}
    \end{pmatrix}\right\}~,
\end{align}

\begin{indentedremark}
Although implicit in equation \eqref{eq:decoupledPNN}, we expect that a complete decoupling of the system is not strictly required in practice and the inputs can simply be projected directly next to one another. Indeed, in practice $\bS$ will usually have some finite `interaction-range'. This will effectively decouple the system into weakly interacting sub-blocks, whose interactions could be further decreased during training by a suitable choice of $\ba_j= \textup{\text{E}}^{(j)} = 0$ for certain subsections, i.e. the input is not projected and the region is not optically probed. We fully expect that our mathematical arguments could be extended from non-interacting to such weakly interacting sub-blocks by a perturbative argument. 
\label{rmk:decoupling}
\end{indentedremark}
\vspace{+1.0em}
Continuing from equation \eqref{eq:decoupledPNN}, we then fix the zeroth (Fourier) component of the output field and define the function
\begin{align}
\label{eq: decoupledPNN_scalar}
    f(\bx) = \mathbf{E}^{\text{out}}(\bx)[0,0]
\end{align}
in the detector plane. For this choice we obtain the following result:
\vspace{-1.0em}
\begin{indentedtheorem}
\label{thm:universality_PNN}
Let $\Omega \subset \R^d$ be a compact domain. For almost all $\bS$, the physical neural network described by equation \eqref{eq: decoupledPNN_scalar} is universal in $C(\Omega)$.
\end{indentedtheorem}
\vspace{1.0em}
\begin{adjustwidth}{3em}{0pt} 
\begin{proof}[Proof of Theorem \ref{thm:universality_PNN}]
We first recall that the 2D discrete Fourier transform of a function $g$ is given by
\begin{align}
    \hat{g}[k,\ell] = \sum_m^{M-1} \sum_n^{N-1} g[m,n] e^{-i2\pi\left(\frac{mk}{M} + \frac{n\ell}{N}\right)}~.
\end{align}
If we now consider the zeroth order  in both dimensions, that is, $\hat{g}[0,0]$, we simply obtain a sum over all components of the function $g$,
\begin{align}
    \hat{g}[0,0] = \sum_m^{M-1} \sum_n^{N-1} g[m,n]~.
\end{align}
From this and equation \eqref{eq:decoupledPNN}, it follows that we can express the zeroth order of $\textbf{E}^{\text{out}}$ as the sum over all components of the vector 
\begin{align}
    \textbf{E}^{\text{out}}[0,0] &= \sum_k \left[\text{r}_\text{m} \big( \I - \text{r}_\text{m}\textbf{T}(\bx) \textbf{S} \textbf{T}(\bx) \big)^{-1} \textbf{E}^\text{in}\right]_k \\
    &= \textbf{e}^{\mathrm{T}} \text{r}_\text{m} \big( \I - \text{r}_\text{m} \textbf{T}(\bx) \textbf{S} \textbf{T}(\bx) \big)^{-1} \textbf{E}^\text{in}~,
    \label{eq:Ezero_zero}
\end{align}
where we have defined $\mathbf{e} = (1,...,1)^{\mathrm{T}}$. Keeping in mind the decoupled form in equation \eqref{eq:decoupledPNN}, equation \eqref{eq:Ezero_zero} can be equivalently written as the sum over components of the individual subsections
\begin{align}
\label{eq:Iout-basic}
    \textbf{E}^{\text{out}}[0,0] = \sum_j \mathbf{e}_{j}^{\mathrm{T}}  \text{r}_\text{m}\big( \I - \text{r}_\text{m}\textbf{T}(\bx) \textbf{S}_j \textbf{T}(\bx) \big)^{-1} \text{E}^{(j)}~,
\end{align}
where we have defined $\mathbf{e}_{j} = (1,...,1)^{\mathrm{T}}$ with the dimensions of $\text{E}^{(j)}$ and the sum runs over the independent subsections. Since the probe beam is a trainable parameter, we can now choose the input probe of each subsection as
\begin{align}
    \text{E}^{(j)} = c_j \mathbf{e}_{j}~,
\end{align}
and obtain
\begin{align}
    \textbf{E}^{\text{out}}[0,0] &= \sum_j c_j \underbrace{\mathbf{e}_{j}^{\mathrm{T}}  \text{r}_\text{m}\big( \I - \text{r}_\text{m}\textbf{T}(\bx) \textbf{S}_j \textbf{T}(\bx)\big)^{-1} \mathbf{e}_{j}}_{=\sigma_j(\bx)} \\
    &= \sum_j c_j \sigma_j(\bx)~.
\end{align}
By encoding the input in the $j^{\,\textup{th}}$ section with scaling and bias parameters, we hence obtain the form
\begin{align}
    \textbf{E}^{\text{out}}[0,0] = \sum_j c_j \sigma_j(\ba_j\bx + \bb_j)~.
\end{align}
Non-degeneracy of the functions $\sigma_j(\ba_j\bx + \bb_j)$ was proven in Theorem \ref{thm:nondegenerate-phase}, and hence, universality follows by applying Theorem \ref{thm:universality_index}.
\end{proof}
\end{adjustwidth}
\vspace{+1.0em}

\indent It is straightforward to extend the proof of Theorem \ref{thm:universality_PNN} to measurement of output components $\textbf{E}^{\text{out}}[k,\ell]$ other than $[0,0]$. In particular, measurement of the output component $[0,0]$ to arbitrary precision is not necessary for universality. We end this section by briefly indicating the required changes in the mathematical argument.\\
\indent First, replacing $\textbf{E}^{\text{out}}[0,0]$ by $\textbf{E}^{\text{out}}[k,\ell]$ results in the following generalization of equation \eqref{eq:Iout-basic}:
\begin{align}
\label{eq:Iout-advanced}
    \textbf{E}^{\text{out}}[k,\ell] = \sum_j \mathbf{\tilde{e}}_j^{\mathrm{T}}  \text{r}_\text{m}\big( \I - \text{r}_\text{m}\textbf{T}(\bx) \textbf{S}_j \textbf{T}(\bx) \big)^{-1} \text{E}^{(j)}~,
\end{align}
where $\mathbf{\tilde{e}}_j$ is a vector whose components are `pure phases', that is, complex numbers of the form $e^{i\phi}$ for suitable angle $\phi$. 

In the original proof of non-degeneracy for the multivariate encoding function $\sigma$, we have the expression [compare with equation \eqref{eq:expansion}],
\begin{align}
\mathbf{e}_j^{\mathrm{T}} \cdot \left[
\bT(\bx)\bS\bT(\bx)
\right]^k
\cdot 
\mathbf{e}_j
= 
\sum_{j_0,j_1,\dots, j_k}
S_{j_0 j_1} \dots S_{j_{k-1} j_k} 
\; 
x_{j_0} x_{j_1}^2 \dots x_{j_{k-1}}^2 x_{j_k}~.
\end{align}
For the more general case described by equation \eqref{eq:Iout-advanced}, we replace $\textbf{e}_j$ by $\mathbf{\tilde{e}}_j$, and we make the choice $\text{E}^{(j)} = \mathbf{\tilde{e}}_j^\ast$ (where $\ast$ denotes the complex conjugate). This then leads to
\begin{align}
(\mathbf{\tilde{e}}_j)^{\mathrm{T}} \cdot \left[
\bT(\bx)\bS\bT(\bx)
\right]^k
\cdot 
\mathbf{\tilde{e}}^\ast_j
= 
\sum_{j_0,j_1,\dots, j_k}
S_{j_0 j_1} \dots S_{j_{k-1} j_k} 
\; 
( e^{i\phi_{j_0}}x_{j_0}) x_{j_1}^2 \dots x_{j_{k-1}}^2 (e^{-i\phi_{j_k}}x_{j_k})~,
\end{align}
As argued above, when taking the derivative $\frac{\partial^{2rd}}{\partial x_1^{2r} \dots \partial x_d^{2r}}
\Big|_{x=0}$ only the terms with $j_0 = j_k$ contribute. For these terms the additional complex phase vanishes. From here, the proof is then identical.
\vspace{-0.5em}
\begin{indentedremark}
    Above we have shown that the proposed system is universal in one output variable. The extension to multiple variables follows quickly, as the system takes the form in equation \eqref{eq:decoupledPNN} in the beam section corresponding to each of the multilenses. Therefore, the system is universal with the number of output variables corresponding to the number of lenses.
\label{rmk:multilens}
\end{indentedremark}
\vspace{-1.0em}
\begin{indentedremark}
    In an experiment we would typically detect intensities, corresponding to the additional transformation $||.||^2$ performed at the output, making it strictly positive. This constrains the learned function to $\R^+$. While this is often acceptable, for some problems functions $f(x)$ with negative outputs might be required. In these cases, we can use the PNN to approximate a function $\tilde{f}(x) = e^{f(x)}$. We can then simply take the logarithm of the result in post-processing. Alternatively, the dataset could be rescaled to ensure that output values are normalized to lie within $[0,1]$, as is commonly done in practice.
\end{indentedremark}

\section{Numerical experiments}

In the main text, we present results of the proposed system on two benchmarks, the MNIST and the Fashion-MNIST. For both tasks, the input consists of $28\times28$ pixel, grayscale images. In our numerical experiments, we downsample these to $14\times14$ pixels, such that the flattened input can be written as a vector $\bx$ with dimension $d=14^2$. The output of our system is a 10-dimensional vector $\textbf{y}$ describing the predicted class of the image. The model architecture we implement performs a function $\textbf{f}(\bx): \R^d \to \R^{10}$ according to
\begin{align}
\label{eq:ML_architecture}
     \text{f}_n(\bx) &= \mathcal{F}\left\{\begin{pmatrix}
    \text{r}_\text{m} \big( \I - \text{r}_\text{m}\textbf{T}(\ba_{n,1}\smallcirc \bx + \bb_{n,1}) \textbf{S}_{n,1} \textbf{T}(\ba_{n,1} \smallcirc \bx + \bb_{n,1}) \big)^{-1} \text{E}^{(n,1)} \\
    \vdots \\
    \text{r}_\text{m} \big( \I - \text{r}_\text{m}\textbf{T}(\ba_{n,r} \smallcirc \bx + \bb_{n,r}) \textbf{S}_{n,r} \textbf{T}(\ba_{n,r} \smallcirc \bx + \bb_{n,r}) \big)^{-1} \text{E}^{(n,r)}
    \end{pmatrix}\right\}[0,0]~,
\end{align}
followed by detection, that is, $||.||^2$. Here $n \in (1,10)$, such that $\text{f}_n$ is the $n^{\textup{th}}$ output variable of $\textbf{f}$(corresponding to each lens), $r$ is the number of input replications per lens and $\textbf{T}$ is a diagonal phase encoding matrix with elements $\text{T}_{lm} = \delta_{lm}e^{ix_m}$, where $\delta_{lm}$ is the Kronecker delta. We have further assumed the reflectivites to be $\text{r}_\text{m}=\text{t}_\text{m}=\frac{1}{2}$. The block-form of equation \eqref{eq:ML_architecture} is physically motivated in Section \ref{sec:linear_recombination}. The trainable parameters of the system are $\ba_{n,j}, \bb_{n,j}$ and $\text{E}^{(n,j)}$. The matrix subblocks $\textbf{S}_{n,j}$ are either randomly intialized and fixed (random $\textbf{S}$) or passed as additional trainable parameters (trained $\textbf{S}$). In both cases, they are enforced to be unitary and symmetric.\\
\indent We implemented the above architecture in Python using PyTorch. For training, we use a cross-entropy loss function and the adam optimizer with autograd. We further use a batch size of $32$ and train the model for $100$ epochs for MNIST and $250$ epochs for the Fashion-MNIST. We used a learning rate of $10^{-3}$ that was gradually decreased during training with a cosine annealing learning rate scheduler. The training dataset for both tasks consisted of 50,000 samples, with 10,000 separate samples set aside for testing. For the Fashion-MNIST, we performed random erasing and random horizontal flips on the training data to prevent overfitting. The results presented in the main manuscript were achieved with these settings, for varying values of $r \in [10,40,90]$ for MNIST and $r \in [10,40,90,160]$ for Fashion-MNIST. These choices correspond to 1, 2, 3, and 4 input replications along each spatial dimension inside each lens region respectively. 
\vspace{-1.0em}
\begin{indentedremark}
    Since we assume $10$ lenses in our system and have an input with $14\times14$ pixels, the largest case ($r=160$) corresponds to the usage of $560\times560$ pixels on the SLM. We note that we assume sufficient spatial separation between the separate copies to achieve a block structure for $\textup{\textbf{S}}$, meaning the effective SLM size would have to be larger to ensure spatial separation of the copies. We note that this assumption is convenient for the proof and to simplify the simulations, but not strictly necessary in a practical system (see \ref{rmk:decoupling}). We assume that similar (if not better) results would be achieved when a completely dense matrix $\textup{\textbf{S}}$ were used for the system as this would provide more degrees of freedom.
\end{indentedremark}

\section{PNNs in the continuous-time formulation with intensity detection}

As discussed in the main text, a continuous-time formulation of equation \eqref{eq:mPNN} can be considered for multivariate PNNs. A key issue that arises is that the intensity detection yields expressions of the form
\begin{align}
    \int_0^T |c_t\sigma (\textbf{a}_t \smallcirc \bx + \bb_t)|^2 dt~,
\end{align}
where an absolute value squared is applied at each point in time. This corresponds to equation \eqref{eq:mPNN} taking the modified form
\begin{align}
\label{eq:bad}
y = \sum_{j} |c_j \sigma(\ba_j \smallcirc x + \bb_j)|^2~.
\end{align}
While this looks similar at first sight, the class of functions in equation \eqref{eq:bad} is not necessarily universal, even if equation \eqref{eq:mPNN} is. \\
\indent A simple counter-example is given for one-dimensional $x\in \R$, and $\sigma(x) = e^{ix}$. In this case, it is well known that Fourier series are dense on any compact interval, and hence 
\begin{align}
\sum_{j} c_j e^{i(a_jx+b_j)}
\end{align}
satisfies universality. In contrast, since $|e^{i(a_j x + b_j)}|^2 \equiv 1$, whatever the choice of $a_j$ and $b_j$, it follows that 
\begin{align}
\sum_{j} |c_j e^{i(a_j x + b_j)} |^2 = \sum_{j} |c_j|^2
\end{align}
can only represent functions that are constant in $x$. Thus, in general, universality can fail dramatically when passing from equation \eqref{eq:mPNN} to equation \eqref{eq:bad}.\\
\indent However, as argued in the main text, we can overcome this issue by including a reference wave to obtain the expression
\begin{align}
    \int |\text{E}_0 + c_t\sigma (\textbf{a}_t \smallcirc \bx + \bb_t)|^2 dt~.
\label{eq:reference_wave}
\end{align}
With rescaling and offset of the output, we can rewrite expression \eqref{eq:reference_wave} as 
\begin{align}
    \omega \int_0^T \left| 1 + c_t \sigma(a_t x + b_t) \right|^2 \, dt + \gamma.
\end{align}
It is for this expression that we can prove the following proposition.
\vspace{-0.5em}
\begin{indentedproposition}
\label{prop:time}
If $\sigma(x)$ is a non-degenerate multivariate nonlinear encoding function, then the class of functions of the form, 
\begin{align}
\label{eq:classy}
\omega \sum_{j} \left| 1 + c_j \sigma(a_j x + b_j) \right|^2 + \gamma
\end{align}
is universal. Such functions can (approximately) be represented in a time-stacked form,
\begin{align}
\label{eq:classint}
\omega \int_0^T \left| 1 + c_t \sigma(a_t x + b_t) \right|^2 \, dt + \gamma~,
\end{align}
for time-varying coefficients $a_t$, $b_t$, and $c_t$. Hence, equation \eqref{eq:classint} also defines a universal class of functions.
\end{indentedproposition}
\begin{adjustwidth}{3em}{0pt} 
\begin{proof}[Proof of Proposition \ref{prop:time}]
Let $f(x)$ be a continuous function that we wish to approximate over a compact domain $D$. Given $\epsilon > 0$, the non-degeneracy of $\sigma(x)$ allows us to find coefficients $a_j, b_j, c_j$, such that 
\begin{align}
\psi(x) = \sum_{j=1}^N c_j \sigma(a_j x + b_j)
\end{align}
approximates $f(x)$ to accuracy $\epsilon/2$:
\begin{align}
\sup_{x \in D}
\left|
f(x) - \psi(x)
\right|
\le \frac{\epsilon}{2}~.
\end{align}
Keeping the coefficients $a_j, b_j, c_j$ fixed, we next note that for any $\delta > 0$, the function
\begin{align}
\psi_\delta(x) = \frac{1}{2\delta}\sum_{j=1}^N \left\{
\left| 1 + \delta c_j \sigma(a_j x + b_j) \right|^2
- 1
\right\}~,
\end{align}
belongs to the class of functions in equation \eqref{eq:classy}.
However, expanding the square, we can write 
\begin{align}
\psi_\delta(x) = \frac{1}{2} \left\{\psi(x) + \psi(x)^\ast\right\}
+ O(\delta)~,
\end{align}
where the $O(\delta)$-term vanishes as $\delta \to 0$, uniformly in $x$. Hence, by choosing $\delta>0$ sufficiently small, we can approximate $\psi(x)$ to arbitrary accuracy by a function in the class described by equation \eqref{eq:classy}. In particular, there exists $\delta > 0$, such that 
\begin{align}
\sup_{x\in D} \left|
\psi(x) - \psi_\delta(x)
\right| 
\le \frac{\epsilon}{2}.
\end{align}
Combining the above estimates, we have shown that for any $\epsilon > 0$, there exists $\psi_\delta$ in the class of functions in equation \eqref{eq:classy}, such that 
\begin{align}
\sup_{x\in D} \left|
f(x) - \psi_\delta(x)
\right|
\le 
\sup_{x\in D} \left|
f(x) - \psi(x)
\right|
+
\sup_{x\in D} \left|
\psi(x) - \psi_\delta(x)
\right|
\le \epsilon. 
\end{align}
\noindent This shows that the class of functions described by equation \eqref{eq:classy} is dense.
\end{proof}
\end{adjustwidth}
\vspace{-1.0em}
\begin{indentedremark}
    We note that the above arguments hold identically for vector-valued coefficients $\textbf{\textup{c}}_j$ and $\textbf{\textup{c}}_t$, as the individual components decouple and can thus be treated separately. From this it follows that the PNN proposed in the main text, given by
    \begin{align}
    \textup{
        $\textbf{f}(\bx, \ba_t, \bb_t, \mathbf{c}_t) = \int \big|\textbf{E}_0 + \mathbf{c}_t \sigma(\mathbf{a}_t \!\smallcirc\! \mathbf{x} + \mathbf{b}_t)\big|^{\smallcirc 2} \, dt,$}
    \label{eq:temporal_detection}
    \end{align}
    with appropriate rescaling in each variable is a universal function approximator.
\end{indentedremark}